
%
%
\input harvmac.tex
%
%
\def\half{{1\over 2}}
\def\T{k_B T}
\def\J{{\cal J}}
\def\Jid{{\cal J}_{id}}
\def\a{_{\alpha}}
\def\b{_{\beta}}
\def\c{_{\gamma}}
\def\d{_{\delta}}
\def\ab{_{\alpha\beta}}

\def\abc{_{\alpha\beta\gamma}}
\def\abcd{_{\alpha\beta\gamma\delta}}
\def\eps{\epsilon}
\def\w{\omega}

\def\n{{\tilde n}}
\def\rrho{{\tilde\rho}}
\def\Rrho{{\vec\rho}}
\def\t{\tau}
\def\Ttau{{\vec\tau}}
\def\Ppsi{{\vec\psi}}
\def\Pphi{{\vec\phi}}

\def\u{{\vec u}}
\def\h{{\vec h}}

\def\H{\hbox{$\cal H$}}
\def\D{\hbox{$\cal D$}}
\def\G{\vec{\vec G}}

\def\S{{\tilde S}}
\def\Z{\hbox{$\cal Z$}}
\def\Zd{\hbox{${{\cal Z}_d}$}}
\def\O{\hbox{$\cal O$}}
\def\V{\vec{\vec V}}
\def\K{\vec{\vec K}}

\def\k{{\vec k}}
\def\r{{\vec r}}
\def\rr{{\vec{\tilde r}}}
\def\R{{\vec R}}
\def\RR{{\vec R}}

\def\z{{\vec\zeta}}

\def\pt{{\partial}}


\centerline{\bf KINETIC THEORY OF FLUX LINE HYDRODYNAMICS:}
\centerline{\bf LIQUID PHASE WITH DISORDER}
\bigskip
\bigskip
\centerline{ Leo Radzihovsky$^{a)}$ and Erwin Frey$^{b)}$ }
\bigskip\centerline{$^{a)}$Lyman Laboratory of Physics}
\centerline{Harvard University}
\centerline{Cambridge, MA 02138}
\bigskip\centerline{$^{b)}$Institut f\"ur Theoretische Physik}
\centerline{Physik-Department der Technischen Universit\"at M\"unchen}
\centerline{James-Franck-Strasse}
\centerline{D--85747 Garching, Germany}


\baselineskip=20pt plus 2pt minus 2pt

\vskip .3in
\centerline{\bf Abstract}
\medskip
We study the Langevin dynamics of flux lines of high--T$_c$ superconductors
in the presence of random quenched pinning. The hydrodynamic theory
for the densities is derived by starting with the microscopic model
for the flux-line liquid. The dynamic functional is expressed as an
expansion in the dynamic order parameter and the corresponding
response field. We treat the model within the Gaussian approximation and
calculate the dynamic structure function in the presence of pinning
disorder. The disorder leads to an additive static peak proportional
to the disorder strength. On length scales larger than the line static
transverse wandering length and at long times, we recover the
hydrodynamic results of simple
frictional diffusion, with interactions additively renormalizing the
relaxational rate. On shorter length and time scales
line internal degrees of freedom significantly
modify the dynamics by generating
wavevector-dependent corrections to the density relaxation rate.

\bigskip
PACS numbers: 64.60C, 05.40, 82.65D \hfill

Submitted to Physical Review B on 05/26/93

\Date{05/93}

\vfill{\eject}

\newsec{Introduction}
\seclab\Intro
\nref\NS{D.~R.~Nelson, {\it Phys.Rev.Lett.}{\bf 60}, 1415
(1988);D.~R.~Nelson and S.~Seung, {\it Phys.Rev.B}{\bf 39}, 9153 (1989);
D.~R.~Nelson, {\it J.Stat.Phys.}{\bf 57}, 511 (1989).}
\nref\HPS{A.~Houghton, R.~A.~Pelcovits and A.~Sudbo, {\it Phys.Rev.B}{\bf 40},
6763 (1989).}
\nref\GSWB{P.~L.~Gammel, L.~F.~Schneemeyer, J.~V.~Waszczak and D.~J.~Bishop,
{\it Phys.Rev.Lett.}{\bf 61}, 1666 (1988).}
\nref\MNi{M.~C.~Marchetti and D.~R.~Nelson, {\it Phys.Rev.B} {\bf 41}, 1910
(1990).}
\nref\GBDKMSW{P.~L.~Gammel, D.~J.~Bishop, G.~J.~Dolan, J.~R.~Kwo, C.~A.~Murray,
L.~F.~Schneemeyer and J.~V.~Waszczak, {\it Phys.Rev.Lett.} {\bf 59}, 2592
(1987).}
\nref\MWYHK{A.~P.~Malozemoff, T.~K.~Worthington, Y.~Yeshurun, F.~Holtzberg and
P.~H.~Kes, {\it Phys.Rev.B} {\bf 38} 7203 (1988).}
\nref\WHF{T.~K.~Worthington, F.~H.~Holzberg and C.~A.~Field, {\it Cryogenics}
{\bf 30}, 417 (1990), and references therein.}

Unlike conventional, low--temperature superconductors
high--T$_c$ superconductors exhibit strong
fluctuations due to the combined effect of the small coherence length $\xi$,
anisotropic layered structure and high temperatures. It has been
argued that the Abrikosov vortex lattice melts as a consequence of these
enhanced thermal fluctuations \refs{\NS-\WHF}.
\nref\CCL{M. Charalambous, J. Chaussy and P. Lejay, {\it Phys. Rev. B}{\bf 45},
509 (1992).}
\nref\SGH{H. Safar, P.L. Gammel, D.A. Huse, D.J. Bishop, J.P. Rice, and D.M.
Ginsberg, {\it Phys. Rev. Lett.} {\bf 69}, 824 (1992).}
\nref\KFW{K.W. Kwok, S. Fleshler, U. Welp, V.M. Vinokur, J. Downey,
G.W. Crabtree and M.M. Miller, {\it Phys. Rev. Lett.} {\bf 69}, 3370 (1992).}
\nref\BNT{E. Brezin, D.R. Nelson, and A. Thiaville, {\it Phys. Rev. B}{\bf 31},
7124 (1985).}
Quite recently there has been experimental evidence \refs{\CCL-\KFW} that
in clean crystal samples (in the absence of twin boundary pinning) the
Abrikosov flux lattice melts via a
first--order phase transition \refs{\BNT}. In the flux liquid state the vortex
lines are free to move through the sample (except for their mutual repulsion)
and will collectively drift in the presence of an external transverse current.
This flux line motion will then in turn generate a finite voltage and lead to
a nonzero linear resistivity.

\nref\FFH{M.~P.~A.~Fisher, {\it Phys.Rev.Lett.} {\bf 62} 1415 (1989);
D.~S.~Fisher, M.~P.~A.~Fisher and D.~Huse, {\it Phys.Rev.B} {\bf 43}, 130
(1990).}
\nref\NV{D.R. Nelson, and V. Vinokur, {\it Phys. Rev. Lett.} {\bf 68}, 2398
(1992); D.R. Nelson and V. Vinokur, Harvard University preprint.}
The phase diagram is changed if one includes the effects of disorder. Depending
on the type and strength of the disorder, the vortex liquid state will persist
down to an irreversibility line associated with a possible second--order
phase transition to
a vortex glass \refs{\FFH}, or Bose glass state \refs{\NV}.
The melting of the flux lattice has clearly very important consequences as
most vividly illustrated in
\fig\PhaseDiagram{(a) Mean-field phase diagram of type II superconductors.
(b) Schematic picture of a phase diagram of high--T$_c$ superconductors which
includes effects of thermal fluctuations and disorder.}
where the mean--field phase diagram
is contrasted with the phase diagram which includes effects of thermal
fluctuations and disorder.

Because the flux-line liquid phase occupies a large portion of the $H-T$ phase
diagram, much of the efforts have been directed toward a better understanding
of the properties of this phase in the presence of disorder.
There has been considerable progress in understanding the static
properties of flux lines in the liquid phase \NS. The resulting
phase is well described by a collection of directed flexible
lines with a line tension related to
$H_{c1}$.
%
%
%
%
The lines traverse the sample in the direction of
the applied magnetic field and at finite temperatures wander throughout
the sample analogously to the Brownian motion executed by atoms or
small molecules in conventional isotropic liquids. This flux-line liquid
in the presence of point disorder is depicted in
\fig\LineLiquid{Schematic picture of flux lines in high--T$_c$ superconductor
in the
presence of pinning disorder (indicated by black circles). Conformation
of the $i$-th line is described by a two-dimensional vector $\vec r_i(z,t)$.}
%
%
%
\nref\GMBGBMK{D.~G.~Grier, C.~A.~Murray, C.~A.~Bolle, P.~L.~Gammel,
D.~J.~Bishop, D.~B.~Mitzi and A.~Kapitulnik, {\it Phys.Rev.Lett.} {\bf 66},
2270 (1991).}
The flux-line states depicted in \PhaseDiagram\
have been extensively studied in many experiments.
Early experiments used the
Bitter technique in which the location of the flux line ends emerging from
the sample is resolved by sprinkling magnetic powder on the sample surface.
\refs{\GBDKMSW,\GMBGBMK}
The disappearance of the regular, hexagonal pattern as the field or
temperature are increased
is suggestive of melting of the Abrikosov flux lattice.
These experiments only directly probe the surface configuration of flux
lines and therefore do not exclude a
possibility of melting confined to the surface. Indeed
theoretical analysis shows
that the surface interaction always dominates in determining
the decay of translational correlations in the asymptotic long-wavelength
limit.
\nref\MNSurface{M.~C.~Marchetti and D.~R.~Nelson, submitted to {\it
Phys.~Rev.~B} (1993).}
\MNSurface\
However, such large length scales have not been probed by the decoration
experiments.
Later vibrating reed experiments by Gammel et al.\GSWB\
have found a signal suggestive of the melting transition.
Although very difficult, more direct measurements of
the bulk properties of the flux-line liquid are possible.
In principal the structure function of the
interacting line liquid can be measured using neutron scattering techniques.
These experiments can directly probe the density correlations in the
line liquid, which are governed by very different physics than
ordinary liquids of point particles with an isotropic structure
function.\nref\CJRF{D.~Cribier, B.~Jacrot, L.~M.~Rao and B.~Farnoux, {\it
Phys.Rev.Lett.} {\bf 9}, 106 (1964).}\CJRF\
Also, recently, new revolutionary electron holographic techniques have
been used to image the motion of flux lines in thin Pb films. This probe,
which can image flux lines in real time, can be used to study mixed
states also in high--T$_c$ superconductors, and can provide envaluable
information about the dynamics of flux lines.
\nref\Hitachi{K.~Harada, T.~Matsuda, J.~Bonevich, M.~Igarashi, S.~Kondo,
G.~Possi, U.~Kawabe, and A.~Tonomura, {\it Nature}, {\bf 360} (1992).}
\refs{\Hitachi}

The interactions between lines, thermal fluctuations and the effects of
disorder are clearly very important and are the main physical features
that must be taken into account by
the theory. Significant progress has been made by mapping the problem of flux
lines onto the quantum statistical mechanics of interacting
$2D$-bosons, where the roles of $\hbar$, temperature and mass of the bosons are
played by the temperature, the inverse sample thickness and line tension,
respectively.
\refs\NS\
With this mapping much of the insight gained from the study of systems like
helium was taken over to the problem of line liquids.
When the lines are long, (i.e. the sample along the applied field direction
is thick), the temperature is high, and the $2D$-density is large, such that
a typical line wandering distance is larger than the average
interline spacing, a highly entangled line liquid results. However,
the entanglement should persist only if the line crossing barriers
are significantly larger than the thermal energy. Although to date
no detailed analysis of line crossing barriers exists, simple estimates give
$U_x/\T_{melt}\sim 10-30$, which translates into very slow relaxational rates.
The entangled phase corresponds to the
Bose-condensed phase in the boson picture.

The static structure function for
the interacting flux lines has been previously computed within the Bogoliubov
approximation taking advantage of the boson mapping \NS.  The contours
of constant scattering intensity form a butterfly pattern with two
peaks, and is quite different from the structure function of isotropic
fluids of particles. At long wavelengths, away from any critical transitions,
the theory
of the vortex liquid can very well be described by a ``hydrodynamic'' model
in terms of density fields, with phenomenological nonlocal coefficients. The
advantage of starting with a microscopic description, however, is
that the long wavelength description of the liquid can be understood in
terms of a more basic, microscopic model, thereby providing a more detailed
understanding.

\nref\MN{M.~C.~Marchetti and D.~R.~Nelson, {\it Phys.Rev.B} {\bf 42}, 9938
(1990);
M.~C.~Marchetti and D.~R.~Nelson, {\it Physica C} {\bf 174},
40 (1991).}
Although the boson mapping has been instrumental for understanding
the static long
wavelength behavior of line-liquids, unfortunately, there does not appear to
be an obvious extension of this mapping to study the real hydrodynamics. Some
progress has been made through phenomenological approaches in which the
dynamic equations of motion are written directly for the coarse-grained
density fields, using the static free energy, with phenomenological nonlocal
coefficients determined by the static structure function. \refs{\MN}

\nref\KSV{Y.~B.~Kim,
M.~J.~Stephen and W.~F.~Vinen, in: Superconductivity vol.~2,
ed. R.~D.~Park (Marcel Dekker, New York, 1969).}
As for the statics
it is useful to obtain the description of the dynamics, starting with
equations of motion for individual interacting flux lines, and
to derive the dynamics for the observable hydrodynamic
quantities like density. The goal is to construct a kinetic theory of
flux-line hydrodynamics analogous to that of point liquids, which was
useful in understanding hydrodynamics of simple liquids many years ago.
With this approach it should be possible to
calculate the hydrodynamic parameters (like line liquid viscosity, if it
exists) which arise completely
from the flux line interaction and entanglement effects and from the single
line microscopic friction. The bare diffusion parameter will be an input to
the theory, and is related to the real microscopic coupling of
flux lines to the underlying crystal
lattice, the Bardeen-Stephen friction coefficient.
\KSV

A question of renormalization of the bare diffusion coefficient
of a tagged line by the presence of the flux-line liquid and the
interaction with this liquid through excluded volume interaction is
also of interest and is related to the line fluid viscosity. It is
expected that if disorder is strong enough and the system is in an
entangled regime a localization phenomena will take place,
driving the renormalized diffusion constant to zero.
\nref\MarchettiDiffusion{See M.~C.~Marchetti, {\it Phys.~Rev.~B}, {\bf 43},8012
(1991) for interesting investigation of flux line interaction renormalization
of the imaginary ``time'' ($i z$) diffusion constant $D_z\equiv T/\eps$.}
\MarchettiDiffusion\

In this paper we take the first step toward a description of the flux-line
liquid in terms of a kinetic theory of line liquids. We introduce a
formalism that is useful for microscopic calculations of the dynamics in the
flux-line liquid phase. The hydrodynamics of the flux-line liquid is studied
by starting with the microscopic description of the interacting flux lines
in terms of the Hamiltonian that includes the repulsive interline
interactions and in the presence of quenched pinning disorder that couples
to the density of lines.  We expect that entanglement effects
are in principle automatically incorporated in the full theory derived
with this kinetic approach.  The repulsive
interaction will inhibit the lines from passing through each
other and for high line densities will result in slowing down of their
dynamics due to these constraints. It is not clear, however, what
simplest approximation to the resulting interacting field theory will retain
these effects.

This paper is organized as follows. In Sec.2 we introduce our
microscopic model for the dynamics of $N$ interacting lines in the
presence of quenched disorder, and in Sec.3 formulate the dynamics in
terms of a more convenient Martin-Siggia-Rose (MSR) description. In Sec.4
the hydrodynamic description is discussed and the microscopic model is
partially recast in terms of the density fields. In Sec.5, by using the
method of auxiliary fields, we integrate out the microscopic degrees of
freedom and derive the effective MSR dynamic functional thereby
obtaining a hydrodynamic description of the interacting flux-line liquid.
In Sec.6 we approximate this theory by truncating the expansion of the
hydrodynamic functional at quadratic order, and within this approximation
calculate the interacting dynamic structure function in the
presence of disorder.  We analyze this dynamic structure function in
various regimes in Sec.7,
and derive the corresponding static structure function demonstrating that
it agrees with the result obtained via the boson mapping method.
We find that on time scales longer than Rouse time (time required for
the single line
excitation of size $L$ to relax elastically) or equivalently
on length scales larger than the transverse line
wandering length, the noninteracting (single flux line)
dynamics is dominated by the center of mass mode with a $k^2$ relaxational
rate. In this regime we find that the {\it interacting} structure function
reduces to that of hydrodynamic frictional diffusion, consistent with
the phenomenological model of Marchetti
and Nelson.\MN\  The interactions between the lines
additively renormalize the
relaxational rate generating a crossover between the noninteracting and
interacting dynamics.
This crossover occurs at flux line length $L=L_I$,
and is physically related to the entanglement length defined in
Ref.~\xref\NS\ .
On wavelengths smaller than the transverse wandering
length and for times shorter than the Rouse time we find
that the noninteracting dynamics is controlled by the internal modes. In this
regime we obtain a complicated wavevector-dependent renormalization of the
dynamics summarized by the interacting structure function.
In Sec.8 we take the phenomenological approach to the hydrodynamic description
of the flux-line liquids and compare with the results of the kinetic approach
derived in Sec.7.
Appendix A describes an independent derivation of the static structure
function using the methods of auxiliary random fields.
In Appendices B and C we derive
the nonlinear terms in the expansion of the hydrodynamic functional and
``hydrodynamic'' Hamiltonian, and analyze the single line dynamics
in various regimes.

\newsec{Dynamical Model}
\seclab\Model
The statistical mechanics of flux-line liquids is very different from
that of a liquid of point vortices because the lines are long and connected.
Compared to these important topological
properties the detailed internal structure of an individual
flux line is relatively unimportant. The essential physics of the
flux-line liquid can therefore be described by the
conformation and position of each line.
These configurations of vortex lines are characterized by a set of $N$
functions ${{\bf R}_i(z)=(\r_i(z),z)}$, where $\r_i(z)$ specifies
the position of
the $i$th line in the $(x,y)$-plane as it wanders along the direction of the
applied magnetic field $\vec H \| \hat z$ through the sample of thickness $L$
(see Fig.2.1). The probability of an equilibrium configuration of $N$
interacting
lines in the presence of disorder is given by the Boltzmann weight
$\exp(-\H/k_BT)$ with
\eqn\Hamiltonian{\H={\epsilon\over 2}\sum_{i=1}^{N}\int_0^L
dz\left(\pt_z \r_i\right)^2+{1\over 2}\sum_{i\neq j=1}^N\int_0^L
dz V(\r_i(z)-\r_j(z))+\sum_{i=1}^N\int_0^L dz U(\r_i(z),z)\;.}
The first term describes the elastic energy of $N$ non-interacting lines with
the line tension $\epsilon$, which for isotropic superconductors is given in
terms of the London penetration length $\lambda$ and the ratio
$\kappa=\lambda/\xi$ by
$\epsilon=(\phi_0/4\pi\lambda)^2\ln\kappa$, with the flux quantum
$\phi_0=h c/2e=2\times10^{-7}$ G cm$^2$ and $\xi$ the superconducting
coherence length and the vortex-line core thickness.  Here we are working in
the
regime for which $\epsilon$ can be approximated by a constant, although with
our formalism we can easily treat the case of a nonlocal elastic energy.

The anisotropic superconductors can be well described by an effective mass
tensor diagonal in the coordinate system with the $z$-axis aligned with
the $\hat{\bf c}$-axis of the crystal,
\eqn\Masstensor{M_{nm}=\left(\matrix{M_1&0&0\cr
                                     0&M_1&0\cr
                                     0&0&M_3\cr}\right)\;.}
\nref\KO{V.~G.~Kogan, {\it Phys. Rev. B}{\bf 24}, 1572 (1981).}
For these anisotropic layered compounds the line tension is considerably
smaller ${\tilde \epsilon} = \epsilon {M_1 \over M_3}$ \refs{\NS},
where $M_1$ is the in--plane anisotropic mass, and $M_3 \approx 10^2 M_1$ is
the much larger effective mass describing the weak Josephson coupling between
the planes. The above formula for ${\tilde \epsilon}$ applies when the flux
lines are dense ($n_0 \lambda_{\perp}^2 \gg 1$, where $\lambda_{\perp}$ is the
in--plane London penetration depth, and $n_0$ the average density of vortex
lines per unit area). In the opposite limit
$n_0 \lambda_{\perp}^2 \ll 1$ the electromagnetic coupling between the planes
is important and one gets ${\tilde \epsilon} = \epsilon / \ln \kappa$
\nref\FisherLosAlamos{D.~S.~Fisher, lectures at The Los Alamos Symposium,
1991.}
\refs{\FisherLosAlamos,\FFH}.

The second term in Eq.\Hamiltonian\ incorporates the flux line interactions.
We treat the
regime in which the line coordinates vary slowly with $z$, although with our
formalism we can easily extend our treatment beyond this regime.
This leads to the interaction energy which can be expressed in terms of a
pair potential which is local in $z$, and in the London limit is given by
\eqn\Vinteraction{V(\r)={\phi_0^2\over8\pi^2\lambda^2}\left[K_0(|\r|/\lambda)
-K_0(|\r|/\xi)\right]\;,}
where $K_0(x)$ is the modified Bessel function with the asymptotics,
\eqn\Klimit{K_0(x)\approx
\cases{ ({\pi\over2x})^{1/2}e^{-x}& for $x\rightarrow\infty$\ ,\cr
         -\ln(x)                  & for $x\rightarrow 0$\ .\cr}}
In Eq.\Vinteraction\ \ we have introduced a short distance cutoff, the
superconducting coherence length $\xi$.

The final term in Eq.\Hamiltonian\ is the contribution of the pinning
impurities to the free energy of the vortex line liquid. For simplicity we
will take the quenched disorder interaction $U(\r)$ to be
Gaussian with zero mean,
\eqna\Ud
$$\eqalignno{\overline{U(\r,z)}&=0\;,&\Ud a\cr
\overline{U(\r_1,z_1)U(\r_2,z_2)}&=F(\r_1-\r_2,z_1-z_2)\;,&\Ud b\cr}$$
where $F(\r_1-\r_2,z_1-z_2)$ encodes the strength and range
of disorder correlations. We will
later specialize to point, line and plane disorder. These three cases appear
to be the most relevant experimentally as we discuss below, with oxygen
vacancies
and interstitials playing the role of point disorder, columnar defects and
grain/twin boundaries as the line and plane disorders.

The model introduced above has been used, in somewhat more
specialized form to describe the static features of the vortex
liquid.\refs\NS\
Nelson et al. used the boson mapping to compute the interacting
static structure function in the dense phase where the density fluctuations
are small and mean field theory is a good description.  Using
renormalization group methods, they were also able to treat
the dilute line liquid near $H_{c1}$, where the fluctuations are strong.
By matching to the dense phase, where the mean-field theory is accurate they
computed the fluctuation-corrected constitutive relation, $B(H)$, and the
static structure function near $H_{c1}$. This work was further extended to
include the effects of point
pinning disorder on the static structure function in this phase.
\nref\NL{D.~R.~Nelson and P.~Le Doussal,
{\it Phys. Rev. B} {\bf 42}, 10113 (1990).}\refs{\NL}

Since we are seeking a hydrodynamic description of the flux line
dynamics we rewrite the interaction and disorder terms in a convenient form,
\eqna\InteractS$$\eqalignno{\sum_{i\neq j}^N\int_0^Ldz
V(\r_i(z,t)-\r_j(z,t))&=\int_{r_1,r_2,z}
V(\r_1-\r_2)\sum_{i\neq j}^N\delta^{(2)}(\r_1-\r_i(z,t))
\delta^{(2)}(\r_2-\r_j(z,t))\;,\qquad&\cr
&=\int_{r_1,r_2,z} V(\r_1-\r_2)n(\r_1,z,t) n(\r_2,z,t)-N L
V(0)\;,\qquad&\InteractS a\cr
\sum_{i=1}^N\int_0^L dz\ U(\r_i(z,t))&=\int d^2r
dz\ U(\r)\sum_{i=1}^N\delta^{(2)}(\r-\r_i(z,t))\;,&\cr
&=\int d^2r dz U(\r)n(\r,z,t)\;.&\InteractS b\cr}$$
In above we defined $\int_{r z}\equiv\int d^2 r dz$.
The self-energy term $N L V(0)$ in Eq.\InteractS{a}, appropriately cutoff by
$\xi$, can be absorbed into the line tension energy, and we will therefore
ignore it in the following analysis.
The form of Eqs.\InteractS\ \ suggests that interactions in the
liquid phase are naturally described in terms of the line density
$n(\r,z,t)=\sum_{i=1}^N\delta^{(2)}\left(\r-\r_i(z,t)\right)$.

\nref\HH{P.~C.~Hohenberg and B.~I.~Halperin, {\it Rev.Mod.Phys.}{\bf 49}, 435
(1977)}
Here we use the model defined by Eqs.\Hamiltonian,\ \Ud\ \ to study the
dynamics of the flux-line liquid. We take
the kinetic theory approach and write down the microscopic dynamic equations
for each of the interacting lines.  We assume that at this basic level the line
fluctuations are overdamped and are therefore governed simply by Model A
type Langevin dynamics \refs{\HH}
\eqn\Langevin{\pt_t \r_i(z,t)=-D{\delta \H\over\delta \r_i(z,t)}+\z_i(z,t)\;.}
For simplicity we assume that the noise $\vec\zeta_i$ is Gaussian
with zero average, and covariance
\eqn\Noise{\langle\zeta_i^a(z,t)\zeta_j^b(z',t')\rangle=
2Dk_BT\delta_{ij}
\delta_{a b}\delta(t-t')\delta(z-z')\;\;\;\;\;,\;\;\;a, b=1,2\;.}

The parameter $D$ is the microscopic kinetic coefficient proportional to
the inverse of the Bardeen-Stephen friction coefficient,\refs{\KSV}
\eqn\BScoeff{\gamma_0={n_0 \pi\hbar^2\over2e^2\xi_{\perp}^2}\sigma_n\;,}
with $\xi_{\perp}$ as the superconducting coherence length in the copper-oxide
planes and $\sigma_n$ is the normal-state conductivity. $D$ describes the
effective drag on a flux line due to the interactions of the normal core
electrons with the underlying solid.
In the simplest case
of center--of--mass--dominated diffusion of noninteracting flux lines
$D$ is proportional to the macroscopic diffusion coefficient (see Sec.7).
The above dynamics of flux lines is quite similar to the dynamics of
polymer melts, with the important difference that for flux lines
there is no solvent.
Since the vortex lines are not moving in any fluid solution, Rouse rather
than Zimm dynamics
\nref\DoiEdwardsdeGennes{M.~Doi and S.~F.~Edwards, {\it The Theory of Polymer
Dynamics}, Oxford University, New York, (1986);
P.~G.~de Gennes, {\it Scaling Concepts in Polymer Physics}, Cornell University,
Ithaca, (1979).}
\DoiEdwardsdeGennes\ applies, and $D$ is a constant.
However, we expect significant changes in the hydrodynamic parameters
of the line liquid coming from the pinning disorder and flux-line
entanglement, when there are significant
barriers to line crossing.

In Eqn.\Langevin\ we have for simplicity ignored the component of the line
drag normal to the flux velocity $\vec v$, which can be accounted for by
adding $\Delta{\hat z}\times\pt_t \r_i(z,t)$ to the left hand side of
Eqn.\Langevin. This term determines the Hall angle $\theta_H$, according to
$\tan\theta_H=\Delta$.\refs{\KSV} and is generally quite small.
\newsec{Martin-Siggia-Rose Formulation of the Model}
\seclab\MSRsection
\nref\BJW{H.~K.~Janssen, {\it Z. Phys.}{\bf B 23}, 377 (1976);
R.~Bausch, H.~K.~Janssen, and H.~Wagner, {\it Z. Phys.}{\bf B 24}, 113 (1976).}
\nref\MSR{P.~C.~Martin, E.~D.~Siggia, and H.~A.~Rose,
{\it Phys. Rev. A} {\bf 8}, 423 (1973)}
We would like to derive a hydrodynamic description of the line liquid
in terms of coarse-grained variables like the density of the flux
lines, starting with the microscopic model presented in Sec.2.
This procedure is most conveniently
implemented using Martin-Siggia-Rose (MSR) formalism \refs{\BJW,\MSR}
which allows the solution to the Langevin equation to be formulated
in terms of a constrained path integral.

\nref\DPJ{C.~De Dominicis, {\it J. Phys. C} {\bf 1}, 247 (1976);
C.~De Dominicis and L.~Peliti, {\it Phys. Rev. B} {\bf 18}, 353
(1978)}\nref\Jensen{R.~V.~Jensen, {\it J.Stat. Phys.} {\bf 25}, 183 (1981).}
The idea is that instead of solving the Langevin equation for the conformation
variables $\r_i(z,t)$ in terms of the random forces $\z_i(z,t)$ and then
computing the correlations functions by averaging over the noise with the
Gaussian weight
\eqn\Wzeta{\omega[\z_i(z,t)]\sim\exp\left[-{1\over4 D k_BT}\int dt\int
dz|\z_i(z,t)|^2\right]}
one can consider $\r_i(z,t)$ as the basic stochastic field with a path
probability density $W[\r_i(z,t)]$ defined by
\eqn\WrWzeta{W[\r_i(z,t)]\D\r_i(z,t)=\omega[\z_i(z,t)]\D\z_i(z,t)}
and eliminate the random forces in favor of the conformation variables. This
is accomplished via a constrained path integral over the noise with the
Langevin equation as the constraint.\refs{\BJW, \DPJ, \Jensen}

To implement the procedure of MSR we note that the noise average of any
observable ${\cal O}[\r_i(z,t)]$, with flux line conformational variables
$\r_i(z,t)$, as the solution of the Eq.\Langevin, can be
expressed in terms of a constrained
path integral,
\eqnn\Oavei$$\eqalignno{\langle
{\cal O}(\r_i)\rangle&=\int{\cal D}\z_i(z,t){\cal D}\r_i(z,t)J[\r_i]
\prod_{a,i,z,t}\delta\left[\pt_t r_i^a(z,t)+D{\delta\H\over\delta
r_i^a(z,t)}-\zeta_i^a(z,t)\right]\cr
&\times{\cal O}(\r_i)\exp\left[-{1\over4Dk_BT}\int dz dt|\z_i(z,t)|^2\right]
\;.&\Oavei\cr}$$
Here $\int{\cal D}\z_i(z,t){\cal D}\r_i(z,t)$ denotes a path integral over
the noise and the conformation of the flux lines with the implied
discretization of
$z$ and $t$ to define the path integral. The quantity $J[\r_i]$ is the
Jacobian of the transformation from $\z_i(z,t)$ to $\r_i(z,t)$ imposed by the
functional $\delta$-function. It ensures that the path probability density,
$W[\r_i(z,t)]$, with
which the averages are computed is still normalized to 1, i.e. $\langle 1
\rangle=1$.

We eliminate the functional $\delta$-function by performing the integral
over $\z_i(z,t)$ and obtain
\eqnn\Oaveii$$\eqalignno{\langle
{\cal O}(\r_i)\rangle&=\int{\cal D}\r_i(z,t)J[\r_i]\cr
&\times{\cal O}(\r_i)\exp\left[-{1\over4Dk_BT}\int dz dt\left(\pt_t r_i^a(z,t)+
D{\delta\H\over\delta r_i^a(z,t)}\right)^2\right]\;.&\Oaveii\cr}$$
Further, it is convenient to perform a Gaussian transformation in order to
``linearize'' the argument in the exponential, the dynamic functional. This is
accomplished by introducing auxiliary fields $\tilde{\r}_i(z,t)$, usually
called
the response fields due their utility in computation of response functions as
will become clear below. (see Appendix C)
\eqn\Oaveiii{\langle
{\cal O}(\r_i)\rangle=\int{\D}\rr_i(z,t)\int{\D}\r_i(z,t)J[\r_i]\
{\cal O}(\r_i)\exp\left(-\Jid[{\rr}_i,\r_i]\right)\;,}
where $\Jid[\rr_i,\r_i]$ is the dynamic functional for a particular
realization of disorder,
\eqnn\Jay$$\eqalignno{\Jid[{\rr}_i,\r_i]&=\int dt\int_0^L
dz\Biggl[{\tilde r}_i^a(z,t)Dk_BT{\tilde r}_i^a(z,t)\cr
&+i{\tilde r}_i^a(z,t)
\left(\pt_t r_i^a(z,t)+D{\delta\H\over\delta
r_i^a(z,t)}\right)\Biggr]\;.&\Jay\cr}$$
It is convenient to work in Fourier representation. We reserve $q$ for
the wavevector in the $z$--direction, $\k$ for the transverse
wavevector and $\w$ for the frequency variable (see below). Since the flux line
ends are freely fluctuating, these boundary conditions are naturally
satisfied by the discrete cosine-Fourier transform with
$\rr_i(z,t)$ and $\r_i(z,t)$ given by,
\eqnn\Fti$$\eqalignno{\r(z,t)&=\int {d\w\over2\pi} \left[\r_o(\w)+
2\sum_{q_n > 0} \cos(q_n z)\, \r(q_n,\w)\right] e^{i\w t}\cr
&=\int_{q_n,\w} e^{i\w t}\cos(q_n z)\, 2^{(\delta_{n,0}-1)}\,
\r(q_n,\w)\;.&\Fti\cr}$$
where $q_n=n\pi/L$ and we have been careful to separately treat the center of
mass,
$q_n=0$, mode. This seperation will turn out to be essential in order to
recover
the correct hydrodynamic result at long wavelengths (see Appendix C).

To examine the explicit form of $\Jid[\rr_i,\r_i]$, we split it up into three
contributions corresponding to the three terms of the Hamiltonian in
Eq.\Hamiltonian,
\eqn\Jsplit{\Jid[\rr_i,\r_i]=\J_0[\rr_i,\r_i]+\J_i[\rr_i,\r_i]+
\J_d[\rr_i,\r_i]\;.}

(i) $\J_0[\rr_i,\r_i]$ is the dynamic functional for the noninteracting flux
lines,
\eqnn\Jo$$\eqalignno{\J_0[\rr_i,\r_i]&=
L\int_{q_n,\omega} \sum_{a,i}
\left[\tilde r_i^a(q_n,\w)Dk_BT{\tilde r}_i^a(q_n,-\w)
+\tilde r_i^a(q_n,\w) \left(\w+i D\eps q_n^2\right)
r_i^a(q_n,-\w)\right]2^{\delta_{n,0}-1}\cr
&={1\over2}\int \RR_i(q_n,\w)\cdot\G(q_n,\w)\cdot\RR_i^T(q_n,-\w)\;,&\Jo\cr}$$
where $\int_{q_n,\k,\w}\equiv\sum_{n}\int d^2 k/(2\pi)^2\int d\w/(2\pi)$, and
we defined,
\eqna\RRG$$\eqalignno{\RR_i(q_n,\w)&=\left(\rr_i(q_n,\w),\r_i(q_n,\w)\right)
\;,&\RRG{a}\cr
\G(q_n,\w)&=L 2^{\delta_{n,0}-1}\left(\matrix{2D\T & \w+iD\epsilon q_n^2\cr
                 -\w+iD\epsilon q_n^2 & 0\cr}\right)\;.&\RRG{b}\cr}$$

(ii) $\J_i[\rr_i,\r_i]$ is the contribution to the total dynamic functional
due to the interaction between flux lines,
\eqn\Ji{\J_i[\rr_i,\r_i]=
- D\sum_{i=1}^N\sum_{j\neq i}\int dt\int_0^L dz \int_k
(\k\cdot\rr_i) V(\k) e^{i\k\cdot(\r_i-\r_j)}\;,}
where $V(\k)$ is the Fourier transformed flux-line interaction,
Eq.\Vinteraction\
\eqna\Ftii$$\eqalignno{V(\k)&=\int d^2r V(\r) e^{-i\k\cdot\r}\;,&\Ftii{a}\cr
&={\phi_0^2\over8\pi^2\lambda^2}{1\over k^2+\lambda^{-2}}\;.&\Ftii{b}\cr}$$

(iii) $\J_d[\rr_i,\r_i]$ is the disorder contribution to the total dynamic
functional, in which we again transformed to Fourier $\k$ space,
\eqn\Jd{\J_d[\rr_i,\r_i]=
- D\sum_{i=1}^N\int dt\int_0^L dz \int_k
(\k\cdot\rr_i) U(\k,z) e^{i\k\cdot\r_i}\;.}

The Jacobian function $J[\r_i]$ in Eq.\Oavei\ depends on the discretization
scheme of
the path integral.  It is simplest to adopt the causal discretization
procedure in which the Jacobian is a constant, independent of $\r_i$ and
can be omitted, provided that simultaneously the ambiguous equal
time correlator is defined to vanish, (see Ref.\xref\Jensen\ for the details).
\eqn\causal{\langle \tilde{r}_i^a(z,t) r_j^b(z',t)\rangle=0\;.}

The averages of physical observables as well as the correlation functions
of these observables can now be expressed in terms of functional derivatives
of the generating
function with respect to external fields that couple to these observables.
For example, $n$-point correlation function of an observable
${\cal O}(\r_i,\rr_i)$ can be obtained from the generating function
${\cal Z}[h(z,t)]$,
\eqnn\Gfun $$\eqalignno
{{\cal Z}[h(z,t)]&=\int{\D}\rr_i(z,t)\int{\D}\r_i(z,t)\cr
&\times\exp\left[-{\J}_{id}[\rr_i,\r_i]+
\int dt \int_0^L dz \ {\cal O}\left(\r_i(z,t),\rr_i(z,t)\right)
h(z,t)\right]\;,\qquad&\Gfun\cr}$$
by functionally differentiating $n$ times with respect to $h(z,t)$,
\eqnn\OOcorr$$\eqalignno
{\Bigl\langle{\cal O}\left(\r_i(z_1,t_1),\rr_i(z_1,t_1)\right)
&\dots{\cal
O}\left(\r_i(z_n,t_n),\rr_i(z_n,t_n)\right)\Bigr\rangle\cr
&={\delta^n {\cal Z}[h(z,t)]\over\delta h(z_1,t_1)\dots\delta
h(z_n,t_n)}\Biggl|_{\{h(z_i,t_i)\}=0}\;.&\OOcorr\cr}$$
Hence, all the information about the dynamics of the flux line liquid
is encoded in the generating functional which we will study in the hydrodynamic
limit, below.
\newsec{Derivation of Hydrodynamic Description}
\seclab\DerivHydro
In the neutron scattering experiments, the neutrons
interact with the magnetic field of the flux lines. The resulting
scattering intensity is therefore proportional to the Fourier transform of the
$2$-point correlation function of the magnetic
field.
\nref\GoldenRule{Because neutrons interact with the magnetic field
only weakly, we assume that the first order perturbation theory, on which the
Fermi-Golden Rule is based, is valid for inelastic neutron scattering from
flux line liquid.}
\nref\qnComment{In this and subsequent sections we have have reverted to
using the continuous Fourier transform for z (defined the same as for $t$).
It turns out that in the large $L$ limit, the breaking of translational
invariance along z-direction and therefore the discreteness of $q_n$ is
only important for $\J_o$ so that $q=0$ is properly taken into account
and is not important for the interactions parts of $\J$.}
\refs{\qnComment}
\refs{\GoldenRule}
The London equation relates the
Fourier transform of the magnetic field components, along and perpendicular
to $\hat z$, to the flux line number and tangent density $n(k,q,t)$,
${\vec\tau}(k,q,t)$, respectively,\refs{\NL,\MN}
\eqna\Brelation$$\eqalignno{&B_z(\k,q,t)={\phi_0\over1+{\lambda}_{\perp}^2
M_1(\k^2+q^2)}\; n(\k,q,t)\;,&\Brelation a\cr
&B_{{\perp}{a}}(\k,q,t)={\phi_0\over1+{\lambda}_{\perp}^2
(M_3 \k^2+M_1 q^2)}\; P^T_{a b}(\k)\tau_b(\k,q,t)\cr
&\;\;\;\;\;\;\;\;\;\;\;\;\;\;\;\;\;\;\;
+ {\phi_0\over1+{\lambda}_{\perp}^2
M_1(\k^2+q^2)}\; P^L_{a b}(\k)\tau_b(\k,q,t)\;,&\Brelation b\cr}$$
where $P^T_{a b}(\k)=\delta_{a b}-k_a k_b/k^2$, $P^L_{a b}(\k)=k_a k_b/k^2$,
and with the densities related to the flux line position,
\eqna\density$$\eqalignno{&n(\r,z,t)
=\sum_{i=1}^N\delta^{(2)}\left(\r-\r_i(z,t)\right)\;,&\density a\cr
&{\vec\tau}(\r,z,t)
=\sum_{i=1}^N{\pt\r_i\over\pt z}
\delta^{(2)}\left(\r-\r_i(z,t)\right)\;,&\density b\cr}$$
Hence inelastic neutron scattering experiments are a direct probe of the
dynamic structure function of interacting flux
lines, i.e. of the flux line density-density correlation function.

\nref\VGLF{V.M. Vinokur, V.B. Geshkenbein, A.I. Larkin, and M.V. Feigel'man,
{\it Sov.Phys JETP} {\bf 73} (3), 610, (1991).}
The transport coefficients like flux-line liquid viscosity and friction
can also be extracted from the dynamic structure function of these
line liquids, and hence can be compared with the experiments that
measure resistance.\MN\
The interacting dynamic structure function also enters
into the perturbative calculation of disorder corrected flux flow velocity in
the high velocity regime.  It has been argued that this relation holds
beyond its expected regime of validity, down to low flow velocities.
The dynamic structure function is therefore directly related to
the current-voltage curves.
\nref\Vinokur{V.~M.~Vinokur,
private communications.}\refs{\Vinokur,\VGLF}

Because it is the flux-line density $n(\r,z,t)$ rather than the microscopic
conformation field $\r_i(z,t)$ that is measured in most of the experiments, it
is more natural to work with the hydrodynamic density variables.
By starting with the microscopic theory defined by $\Jid[\rr_i,\r_i]$,
Eqs.\Jsplit-\Jd, and integrating out the
microscopic conformational degrees of freedom $\r_i(z,t)$ and $\rr_i(z,t)$,
with the constraint that the densities $n(\r,z,t)$ and corresponding response
field $\n(\r,z,t)$ remain fixed, we derive the dynamic functional of the
density fields. With this new form of the dynamic theory the density
correlation and response functions are easily computable and the
approximations that are necessary can be more easily physically motivated,
because the density and its correlations are directly observable.

We begin by introducing a new density response field, $\n(\r,z,t)$,
\eqn\densityi{\n(\k,z,t)=\T D\sum_{i=1}^N
\left[ \k\cdot\rr_i (z,t) \right] \; e^{-i\k\cdot\r_i (z,t)}\;,}
in addition to the physical density field $n(\r,z,t)$ already introduced
in Eq.{\density a}. The new response field $\n(\r,z,t)$ will earn its name
by generating dynamic density response functions (see below and
Appendix C). Here we
will only study correlation and response functions of the number density
field and therefore will trace over the tangent density
field ${\vec\tau}(\r,z,t)$ that is related to the fluctuations
of the magnetic field in the $ab$-plane, Eq.{\density b}.

We reexpress the interaction parts of the dynamic functional,
Eqs.{\Ji,\ \Jd}, in terms of the density fields,
\eqna\JJi$$\eqalignno{\J_i[\rr_i,\r_i]&=i D\int_{t,z}
\int_{\r,\r'}\sum_{i=1}^N\delta^{(2)}(\r-\r_i(z,t))
\sum_{j=1}^N\rr_j\cdot{\pt\over\pt\r_j}\delta^{(2)}(\r\ '-\r_j(z,t))
V(\r-\r\ ')\;,\cr
&\;&\JJi a\cr
&={1\over\T}\int_{t,z}\int_{\r} \int_{\r'} n(\r,z,t)\n(\r\ ',z,t)
V(\r-\r\ ')\;,&\JJi b\cr}$$
and the disorder contribution,
\eqna\JJd$$\eqalignno{\J_d[\rr_i,\r_i]&=i D
\int_{t,z,\r}\sum_{i=1}^N\rr_i\cdot{\pt\over\pt\r_i}\delta^{(2)}(\r-\r_i(z,t))
U(\r,z)\;,&\JJd a\cr
&={1\over\T}\int_{t,z,\r}\n(\r,z,t)U(\r,z)\;.&\JJd b\cr}$$
It is important to remember that in the above equations $\n(\r,z,t)$ and
$n(\r,z,t)$ are not fields independent of the fundamental microscopic fields
$\rr_i(z,t)$ and $\r_i(z,t)$ and at this point are only a notational
convenience.

Although the mean-field theory of the flux-line liquid hydrodynamics is
described by the average densities $n_0=\langle n(\r,z,t)\rangle$
and $\n_0=\langle \n(\r,z,t)\rangle$, it is
the fluctuations about the mean density that are nontrivial and
are probed by the scattering experiments. We therefore introduce fields,
\eqna\densityii$$\eqalignno{\rho(\r,z,t)&=n(\r,z,t)-n_0\;,&\densityii a\cr
\rrho(\r,z,t)&=\n(\r,z,t)-\n_0\;.&\densityii b\cr}$$
describing the deviations from the average uniform densities
$n_0=\langle n(\r,z,t)\rangle$ and $\n_0=\langle \n(\r,z,t)\rangle$.

It is convenient to work in the canonical ensemble in which the number of flux
lines $N$ is fixed (i.e. total magnetic field through the sample is constant),
and for fixed sample area $A$, $n_0=N/A$.
Substituting Eqs.\densityii{a,b} into Eqs.\JJi \ , \JJd \ , using the
conservation of flux lines identities,
\eqna\conserve$$\eqalignno{\int_{\r}
\rho(\r,z,t)&=\rho(\k=0,z,t)=0\;,&\conserve
a\cr
\int_{\r} \rrho(\r,z,t)&=\rrho(\k=0,z,t)=0\;,&\conserve b\cr}$$
and ignoring the remaining constant terms that do not effect the dynamics, we
obtain interaction part of dynamic functional in Fourier space,
\eqna\JJJi$$\eqalignno{\J_i[\rr_i,\r_i]&={1\over\T}\int_{\k,q,\w}
\rrho(\k,q,\w)V(\k)\rho(-\k,-q,-\w)\;,&\JJJi a\cr
&={1\over 2\T}\int_{\k,q,\w} \rho\a(\k,q,\w)V\ab(\k)\rho\b(-\k,-q,-\w)
\;,&\JJJi b\cr}$$
where we define,
\eqna\JJidef$$\eqalignno{\Rrho&=\left(\rrho,\rho\right)\;,&\JJidef a\cr
\V&=\left(\matrix{0& V(\k)\cr
                  V(\k)&0\cr}\right)\;,&\JJidef b\cr}$$
For the disorder contribution to the dynamic functional we obtain,
\eqna\JJJd$$\eqalignno{\J_d[\rr_i,\r_i]&={1\over\T}\int_{\k,q,\w}
\rrho(\k,q,\w)U(-\k,-q)\delta(\w)\;,&\JJJd a\cr
&={1\over\T}\int_{\k,q,\w}
\rho\a(\k,q,\w) u\a(-\k,-q,-\w)\;,&\JJJd b\cr}$$
where we defined disorder vector,
\eqn\Uvec{\u(\k,q,\w)=\left(U(\k,q,\w)\delta(\w),0\right)\; .}

As discussed in Sec.3, we construct the dynamic generating functional
for the computation of density
correlation and response function,
by choosing the operator $\O\a(\rr_i,\r_i)=\rho\a(\r,z,t)$ and coupling
an external field $h\a(\r,z,t)$ to this density observable,
\eqn\Gfuni{\Zd[\h(\r,z,t)]=\int{\D}\RR\
\exp\left[-\Jid[\RR(z,t)]+\int_{\r,z,t}\Rrho(\r,z,t)\cdot\h(\r,z,t)\right]\; .}
In the dynamic functional, expressed in terms of densities, the disorder
field, that couples linearly to the density field, can be
integrated out exactly.
It is important to note that in the dynamic formulation presented here we
can integrate over the quenched disorder directly at the level of the dynamic
generating functional $\Zd[h]$. This is to be contrasted with static
calculations where one must first compute the physical observables
like free energy or correlation functions for particular realization of
disorder and then perform the average over the disorder.(see Appendix A)
This leads to
the usual problems of disorder-averaging $\ln Z_d$ or $Z^{-1}_d$ (since the
static averages have to be normalized by the partition function $Z_d$),
problem usually handled using the ``replica trick''.  In the dynamic
calculations presented here no such complications arise because the
``dynamic partition function'' $\Zd[h=0]=1$, a constraint enforced
by the MSR Jacobian in
Eq.\Oavei\  . We then integrate out the quenched disorder exactly, using
the assumption of Gaussian disorder, Eq.\Ud\ .
Upon averaging $\Zd[h]$ over the quenched pinning potential $\u$, we obtain
\eqna\Zquench$$\eqalignno{\Z[\h(\r,z,t)]&=\overline{\Zd[\h]}
\;,&\Zquench{a}\cr
&=\int{\D}\RR\
\exp\left[-\J[\RR(z,t)]+\int_{\r,z,t}\Rrho(\r,z,t)\cdot\h(\r,z,t)\right]
\;,\qquad\qquad&\Zquench{b}\cr}$$
where,
\eqn\Jquench{\J[\RR(z,t)]=\J_o[\RR]+\J_{int}[\Rrho]\;,}
and
\eqn\Jquenchi{\J_{int}[\Rrho]=
{1\over 2\T}\int_{\k,q,\w} \rho\a(\k,q,\w)K\ab(\k,q,\w)\rho\b(-\k,-q,-\w)\;,}
with
\eqn\Kmatrix{\K(\k,q,\w)=
\left(\matrix{-F(\k,q)\delta(\w)/\T&V(\k)\cr
                 V(\k)&0\cr}\right)\;\;.}

The interaction and disorder contributions have now been expressed as a
quadratic function of the density fields $\rho\a$. However, the
computation of the dynamic generating function $\Z[\h]$ is still
nontrivial because (i) the independent variables are
$\RR$ and not $\rho\a$, and (ii) the noninteracting part of the dynamic
functional $\J_0[\RR]$ cannot be trivially rewritten in terms of the
density fields. To make progress however we can proceed with an uncontrolled
variational approximation. We replace
$\J_0[\RR]$ by an Ansatz that is Gaussian in the density fields
\eqn\Ansatz{\J_0[\RR]\rightarrow\J_0[\rho\a]=
{1\over 2 n_0}\int_{\k,q,\w}
\rho\a(\k,q,\w)\Gamma^{-1}\ab(\k,q,\w)\rho\b(-\k,-q,-\w)\;,}
and use a measure $\D\rho\a$ instead of $\D\RR$ in the functional
integral. The matrix $\Gamma\ab$ is determined by requiring that the
correlation functions of the Gaussian Ansatz agree with the original
noninteracting theory where the averages are performed with measure
$\D\RR\exp(-\J_0[\RR])$. The simplest requirement is that the structure
functions in two theories agree. This completely fixes $\Gamma\ab=S^0\ab$,
where $S^0\ab$ is the structure function for a single flux line. Combining
with $\J_{int}$ this approximation then gives the full dynamic functional,
\eqn\FullAnsatz{\J[\rho\a]=
{1\over 2\T n_0}\int_{\k,q,\w}
\rho\a(\k,q,\w)\left[n_0 K\ab(\k,q,\w)+
\T{S^0\ab}^{-1}(\k,q,\w)\right]\rho\b(-\k,-q,-\w)\;,}
from which all the dynamic correlation and response functions of
density fields can be computed.

In the next section we will treat the full
dynamic functional more rigorously.
We will show how the above ad hoc approximation emerges as a result of
truncation of a systematic expansion
of the dynamic functional in the density fields.

\newsec{Decoupling Flux Line Dynamics (Method of Auxiliary Random Fields)}
\seclab\ARF
As we have already noted the density fields $\rrho(\r,z,t)$ and
$\rho(\r,z,t)$ are not independent of the fundamental microscopic fields
$\rr_i(z,t)$ and $\r_i(z,t)$ and at this point are only a notational
convenience.
The computation of the generating function in the above equation is made
nontrivial by the highly nonlinear flux line interactions $V(\r)$, when
expressed in terms of the kinetic fields $\rr_i(z,t)$ and $\r_i(z,t)$.
To obtain interaction corrections to the noninteracting lines dynamics via a
straight perturbation theory in powers of $V(\r)$ is possible but is a
nontrivial
exercise. \nref\SeungThesis{S.~Seung, {\it Statistical Mechanics of Lines and
Surfaces}, Ph.D. Thesis, Harvard, 1990}\refs{\SeungThesis}
We note however that this interaction is simply quadratic when expressed  in
terms of the density fields. The idea is then to transform the functional
integral from the microscopic variables to independent density fields. This
can be accomplished using the method of auxiliary random fields
that has been previously applied to treat both statics and dynamics
of polymers and is described below.
\nref\Method{T.~Ohta and A.~Nakanishi, {\it J. Phys. A} {\bf 16}, 4155
(1983); A.~Nakanishi and T.~Ohta, {\it J. Phys. A} {\bf 18}, 127 (1985).
\hfil\break
For applications in statics see:\hfil\break
G.~H.~Fredrickson and E.~Helfand, {\it J. Chem. Phys.} {\bf 87} (1), 697
(1987);\hfil\break
S.~M.~Bhattacharjee and J.~J.~Rajasekaran, {\it Phys. Rev. A} {\bf 44}, 6202
(1991).\hfil\break
For applications to dynamics see:\hfil\break
P.~R.~Baldwin, {\it Phys. Rev. A} {\bf 34}, 2234 (1986);\hfil\break
G.~H.~Fredrickson and E.~Helfand, {\it J. Chem. Phys.} {\bf 93} (3), 2048,
(1990).}\refs\Method

We introduce a set of transformations that transform the functional of
$\R(\r,z,t)$
into a functional for the density fields. We accomplish this by introducing
an independent auxiliary density fields $\Ppsi(\r,z,t)$ constrained by the
functional $\delta$-function to equal to the physical density fields
$\Rrho(\r,z,t)$ for all $\r,z,t$,
\eqna\si $$\eqalignno{{\cal Z}[h(\r,z,t)]&=\int{\cal D}\vec R\;
e^{-\J_o[\R]-\J_{int}[\Rrho]+\int\Rrho\cdot\h}\int{\cal
D}\vec\psi\prod_{\r,z,t}\delta(\vec\psi(\r,z,t)-\Rrho(\r,z,t))\;\;\;
\qquad\qquad &\si{a}\cr
&=\int{\cal D}\vec\psi\; e^{-\J_{int}[\Ppsi]+\int\vec\psi\cdot\h} \left\langle
\delta(\vec\psi-\Rrho)\right\rangle_o\;,&\si{b}\cr}$$
where the average is performed with the Gaussian dynamic measure
$e^{-\J_o[\R]}$ of $N$
noninteracting flux lines, with $\J_o[\R]$ given by Eq.\Jo.  We observe that
with this transformation the troublesome interaction is indeed
quadratic in $\Ppsi(\r,z,t)$ and therefore in this sense can be treated
nonperturbatively.
The nontrivial part of the calculation reduces to the
computation of the average
of the functional $\delta$-function with the Gaussian measure
$e^{-\J_o[\R]}$.

Using the functional representation of the $\delta$-function we obtain,
\eqna\delfunc $$\eqalignno{\left\langle\delta(\vec\psi-\Rrho)\right\rangle_o
&=\left\langle\int{\cal D}\vec\phi
e^{i\int\Pphi\cdot(\Ppsi-\Rrho)}\right\rangle_o &\delfunc{a}\cr
&=\int{\cal D}\vec\phi e^{i\int\Pphi\cdot\Ppsi}\left\langle
e^{-i\int\Pphi\cdot\Rrho}\right\rangle_o\;. &\delfunc{b}\cr}$$
Noting that $\Rrho(\r,z,t)=\sum_{i=1}^N\Rrho_i(\r,z,t)$, with
$\Rrho_i(\r,z,t)$ as the single line density, and since
the average for each line
is identical, Eq.\delfunc\  further reduces to,
\eqn\delfunci{\left\langle\delta(\vec\psi-\Rrho)\right\rangle_o
=\int{\cal D}\vec\phi e^{i\int\Pphi\cdot\Ppsi}\left[\left\langle
e^{-i\int\Pphi\cdot\Rrho_1}\right\rangle_o\right]^N\;.}

Combining this result with interaction contribution to the hydrodynamic
functional in Eq.\si\,  we obtain the dynamic generating function
\eqn\Zii{{\cal Z}[\vec h]=\int{\cal
D}\vec\psi\int{\cal D}\vec\phi\exp\left[-\J[\Ppsi,\Pphi]
+\int\Ppsi(\r,z,t)\cdot\vec h(\r,z,t)\right]\; ,}
in terms of the hydrodynamic functional $\J[\Ppsi,\Pphi]$,
\eqn\Jtot{\J[\Ppsi,\Pphi]=\J_{int}[\Ppsi]+n_o\Gamma[\Pphi]
-i\int\Pphi\cdot\Ppsi\;,}
with the new contribution $n_o\Gamma[\Pphi]$ coming from the average of the
$\delta$-function in Eq.\delfunci\ and is expressed in terms of a single
flux line cumulant expansion,
\eqnn\Geff
$$\eqalignno{
 -\Gamma[\Pphi]=
 &{(-i)^2\over 2!}\int\phi\a\phi\b\Gamma^{(2)}\ab+{(-i)^3\over 3!}
  \int\phi\a\phi\b\phi\c\Gamma^{(3)}\abc \cr
+&{(-i)^4\over 4!}
 \int\phi\a\phi\b\phi\c\phi\d\Gamma^{(4)}\abcd+\ldots\;,
 &\Geff\cr}$$
where $\Gamma^{(m)}_{\alpha_1 \alpha_2 \ldots\alpha_m}$ is a connected
correlation function of a single line density fluctuations $\Rrho_1[\R_1]$,
and in coordinate space $x=(\r,z,t)$ is
\eqn\Gammas{\Gamma^{(m)}_{\alpha_1 \alpha_2 \ldots\alpha_m}(x_1,x_2,\ldots,x_m)
=A\langle\rho_{\alpha_1}(x_1)\rho_{\alpha_2}(x_2)\ldots\rho_{\alpha_m}(x_m)
\rangle_o^{c}\;,}
where the area factor $A$ was inserted for convenience. In above we have
also dropped the single line label $1$ on $\Rrho_1$ since for a
while we will be dealing with a single line density and no confusion should
arise.  The cumulant functions
$\Gamma^{(m)}_{\alpha_1 \alpha_2 \ldots\alpha_m}(x_1,x_2,\ldots,x_m)$ can be
systematically computed for any $m$ (see Appendices B and C).

We note that the density field $\Ppsi$ appears only
linearly and quadratically in Eq.\Zii\ \ and therefore can be integrated out
exactly. This leads to the dynamic
functional that is expressed not in terms of the physical densities
$\Ppsi$ but in terms of the auxiliary fields $\Pphi$, conjugate to
$\Ppsi$.  Since the hydrodynamic functional is derived in the
presence of sources $\vec h$ the density correlation functions and
response functions of $\Ppsi$ can still be easily computed as functional
derivatives with respect to these sources (see Eq.\OOcorr).
Upon integrating out $\vec\psi$ we obtain,
\eqn\Ziii{{\cal Z}[\vec h]=\int{\cal
D}\vec\phi e^{-\J[\Pphi,\h]}.}
where,
\eqn\Jayii{\J[\vec\phi,\h]=n_o\Gamma[\Pphi]-{\T\over2}\int
(i\phi\a+h\a)K\ab^{-1}(i\phi\b+h\b)\; .}

The above equations define the hydrodynamic field theory of flux lines
derived directly from the microscopic interacting single line dynamics.
No approximations have been made up to now, and both the flux line
interactions and interaction with disorder have been treated exactly.
To fully solve the hydrodynamic theory we need to compute the generating
function ${\cal Z}[\vec h]$ by integrating over $\Pphi$.  Since the resulting
theory has a form of an interacting field theory, it cannot
be solved exactly. However, one could try to treat the nonlinear
interactions perturbatively with various techniques.
In the long wavelength limit (which can in
principle be studied with sufficiently
low angle neutron scattering experiments) the $\Pphi$
fluctuations can be treated with renormalization group methods
as was done for the statics of polymers solutions by Ohta and Nakanishi\Method\
{}.
This procedure would lead to renormalized vertex functions,
$\Gamma^{(m)}$, corrected by the thermal fluctuations of $\Pphi$. As will be
shown in the next section, some of the flux line fluctuations and
interactions are nontrivially taken into account even if we truncate the
hydrodynamic functional at the quadratic order.  The resulting truncated
theory can be solved exactly, allowing calculations of
dynamic and static properties of the interacting flux-line liquid.

\newsec{Gaussian Approximation to the Dynamic Functional}

In the previous section we derived the expression for the dynamic
generating functional. It is expressed in terms of an infinite series
of interactions in the auxiliary field $\vec\phi$ and in
principle allows for calculation of any hydrodynamic correlation
function of the flux-line liquid. Near $H_{c1}$, where the magnetic field is
weak and the flux-line liquid is dilute, the fluctuations in the density
fields are large, i.e. comparable to the average line density. In this
critical region the nonlinear interactions must be carefully taken into
account. Using renormalization group methods the dynamics of dilute
line-liquids can be studied, as was done for the statics. \NS\
Similarly, near $H_{c2}$, where the flux-line liquid is dense, the vortex
cores begin to overlap,
the effects of nonlinear interactions are large
and renormalization group treatment is again needed to take the
nonlinearities into account.
\nref\HcComment{Near $H_{c2}$ the {\it amplitude} of the BCS order parameter
experiences large fluctuations that must be carefully taken into
account, see Ref.\xref\BNT . Also in this regime the line crossing
barriers are very small.}\HcComment\
Here, however, we study the dynamics of a
semidilute flux-line liquid, away from
such critical regions, i.e. in the regime where $H_{c1} << H << H_{c2}$.
In this regime $n_0$ is
large, hence the method of steepest descents applied to Eqs.\Ziii,\Jayii\
allows us to treat ${\cal Z}[\vec h]$ in mean-field theory. The
nonlinear interactions can then be treated in perturbation theory, but in this
semidilute regime we do not expect qualitative corrections to the results
of our mean-field approximation.  In this section we will therefore
use a Gaussian approximation for computation of $\Z[\h]$ by
truncating expansion of the dynamic functional Eq.\Geff\ ,
at quadratic order in $\Pphi$.

The remaining $\Gamma^{(2)}\ab(\k,q,\w)$ vertex function has a clear physical
interpretation of a matrix whose $(2,2)$ and $(2,1)$ components are
the noninteracting dynamic structure function $S^o(\k,q,\w)$ and response
function
$\S^o(\k,q,\w)$ of the flux line liquid, as can be seen from Eq.\Gammas\ .
$\Gamma^{(2)}\ab(\k,q,\w)$ is computed and analyzed in various regimes
in Appendix C.
With the truncation at quadratic order we obtain the interacting stucture
and response functions $S\ab(\k,q,\w)$ in terms of the noninteracting
ones
$S^o\ab(\k,q,\w)=\Gamma^{(2)}\ab(\k,q,\w)$ and therefore will be able
to study the effects of flux-line interactions and quenched disorder on the
dynamics of the flux-line liquid.

In the Gaussian approximation it is more convenient to
return to the Eqn.\Zii\ \ and integrate out
the $\Pphi$ field, thereby producing a hydrodynamic functional $\J[\Ppsi]$
expressed directly in terms of physical density fields $\Ppsi$. Upon performing
all the calculations in Fourier space ($q$ and $\k$ are reserved for
wavevectors in the $z$ and the transverse directions, respectively) we obtain,
\eqnn\Ziv
$$\eqalignno{
 {\cal Z}[\vec h]=
  \int{\cal D}\vec\psi \;
  \exp \Biggl\{
  \int_{\k,q,\w} \Biggl( -{1\over2} &\psi\a(\k,q,\w)
  {\hat K}\ab(\k,q,\w)\psi\b(-\k,-q,-\w)+ \cr
  +&\psi\a(\k,q,\w)
  h\a(-\k,-q,-\w) \Biggr) \Biggr\} \;. &\Ziv\cr  }
$$
where,
\eqn\Kmatrix{{\hat K}\ab(\k,q,\w)={1\over\T}K\ab(\k,q,\w)+{1\over
n_o}{(S^o)}^{-1}\ab(\k,q,\w)\; .}
An integration over $\Ppsi$ leads to the dynamic generating function
within the Gaussian approximation,
\eqn\Ziv{{\cal Z}[\h]=
\exp\left\{{1\over2}\int_{\k,q,w}h\a(\k,q,\w)
{\hat K}^{-1}\ab(\k,q,\w)h\b(-\k,-q,-\w)\right\}\;.}
Functionally differentiating twice with respect to external field
$\h(\k,q,\w)$, we obtain the
interacting correlation/response function matrix
$S\ab(\k,q,\w)={\hat K}^{-1}\ab(\k,q,\w)$,
\eqn\Str{S\ab(\k,q,\w)=\left(\matrix{0&\S(\k,q,\w)\cr
                  \S(-\k,-q,-\w)&S(\k,q,\w)\cr}\right)\qquad,}
where $S_{2 2}=S(\k,q,\w)$ and $S_{1 2}=\S(\k,q,\w)$ are the interacting
hydrodynamic structure and response functions, respectively, for
the flux line liquid in the presence of disorder,
\eqna\Struct $$\eqalignno{S(\k,q,\w)&=
{F(\k,q)\delta(\w)|\S^o(\k,q,\w)|^2 n_o^2/(\T)^2+
n_o S^o(\k,q,\w)\over|V(\k) \S^o(\k,q,\w)n_o/\T+1|^2}\;,&\Struct a\cr
&&\cr
\S(\k,q,\w)&=
{V(\k) n_o^2 |\S^o(\k,q,\w)|^2/\T+n_o \S^o(\k,q,\w)
\over|V(\k) \S^o(\k,q,\w) n_o/\T+1|^2}\;.&\Struct b\cr}$$
Note that in Eq.\Str\  the $S_{1 1}$ component
vanishes, $\langle{\tilde\psi}(\r,z,t'){\tilde\psi}(\r\ ',z',t')\rangle=0$.
This is a consequence of the fluctuation dissipation theorem
and causality, and implies that dynamics of the density field $\psi_2$
encoded in the effective dynamic functional in Eq.\Ziv\ can equivalently be
described by a linear differential equation for the density field $\Ppsi$.

\newsec{Details of the Dynamic Structure Function}
\seclab\Details
In this section we will analyze the dynamic structure
function $S(\k,q,\w)$. Information about the behavior of
$\S(\k,q,\w)$ can be obtained from $S(\k,q,\w)$ by using
Fluctuation Dissipation Theorem (FDT),
\eqn\FDTtext{\pt_t\Gamma^{(2)}_{2 2}(\k,z,t)=
\Gamma^{(2)}_{1 2}(\k,z,t)-
\Gamma^{(2)}_{1 2}(\k,z,-t) \;.}
The interacting dynamic structure function in Eq.\Struct{a}
is expressed in terms of
the noninteracting structure and response functions, $S^o(\k,q,\w),
\S^o(\k,q,\w)$ of the vortex liquid and depends on the correlation function of
the disorder $F(\k,q)$ as well as the interline interaction $V(\k)$.

The details of a single line dynamics are presented in Appendix C.
There are two regimes with very different dynamic behavior
that are seperated by a characteristic time, $t_{Rouse}=L^2/(D\eps)$.
The Rouse time is the time required for the center of mass of the flux line to
diffuse its transverse radius of gyration. Equivalently, for times
larger than $t_{Rouse}$ the diffusion of the center of mass dominates over the
internal mode dynamics. The dynamics of a single flux line
for $t\gg t_{Rouse}$ and $t\ll t_{Rouse}$ is depicted in
\fig\Rouse{2D--projected configuration of a diffusing flux line
for (a) $t\gg t_{Rouse}$ and (b) $t\ll t_{Rouse}$.
In (a) the progression in time is illustrated, with the inset
showing a snap-shot configuration, ``diffusion'' in $z$--direction.
For $t\gg t_{Rouse}$ the flux line diffusion is
dominated by the center--of--mass mode and the dynamics is that of
a rigid rod. The dashed line shows the trajectory of the center of mass.
In (b) the flux line appears frozen with dynamic
fluctuations (due to the center of mass and internal modes)
small relative to its transverse size. The inset show evolution in time.}.

We find that for times larger than the Rouse time, $t_{Rouse}=L^2/(D\eps)$ the
diffusion is dominated by the center of mass mode. In this case the average
transverse distance diffused is simply
\eqn\CMDiff
{r^{cm}_D(t)=
\sqrt{ {1 \over 2} \langle \left[  \vec r_0(t) - \vec r_0 (0) \right]^2 \rangle
}
=\sqrt{2 D \T t \over  L} \propto t^{1/2} \; .}
We note that
the flux line diffusion coefficient $D\T/L$ is scaled down by $L$ with
respect to the point vortex diffusion constant $D\T$ in the $ab$-plane.
This $1/L$ behavior has
been previously derived in the appendix of Ref.~\xref\MN\ in terms of a simple
model of point vortices coupled in the $z$--direction.

In the opposite limit, $t< t_{Rouse}$, the flux line diffusion is dominated by
the internal modes. For two points on the flux line separated by
a distance $z\ll (D\eps t)^{1/2}$ the auto--correlation function becomes
indendent of $z$. This allows us again to extract a transverse
diffusion distance
\eqn\IMDiff
{r^i_D(t)=
\sqrt{ {1 \over 2} \langle \left[  \vec r(z,t) - \vec r (0,0) \right]^2
\rangle}
\approx \sqrt{ 2 \T \left( {D \over \pi \eps} t \right)^{1/2} }
\propto t^{1/4} \; , }
but now with an anomalous $t$--dependence.
%
%
%
%
%
This behavior can be understood
as two ``random walks'' on top of each other; the flux line segment executes
a random walk on the flux line conformation, which can be thought of being
generated by a random walk (fictitious dynamics along the $z$-axis).
For $z \gg (D\eps t)^{1/2}$ one gets ``diffusion'' in the time--like variable
$z$.
The corresponding square root of the mean square displacement $\sqrt{2\T
|z|\over\eps}$
corresponds to the projected 2D radius of gyration of a flux line of length
$|z|$.
At intermediate scales $z \geq (D\eps t)^{1/2}$ the $z,t$-dependence of the
segment--correlation
function has to be taken into account, and one finds
\eqn\IMFull
{\langle \left[  \vec r(z,t) - \vec r (0,0) \right]^2 \rangle
= {4 \T \over \eps} |z| f \left( {D \eps |t| \over z^2} \right)
\; ,}
with $f(x)$ shown in
\fig\ImodesFig{Scaling function $f(x)$ describing
single flux line diffusion due to
internal modes, displayed on a double logarithmic scale.}.

The dynamic crossover described above can be equivalently extracted from the
behavior of the $\w$-pole in the noninteracting structure function. For
${\T \over \eps} \, k^2 L \ll 1$ the center of mass diffusion dominates
the hydrodynamics and the noninteracting relaxation rate is
$\Gamma_{cm}(\k)=D\T k^2/L$. In the opposite limit of
${\T \over \eps} \, k^2 L \gg 1$ the internal modes
dominate the dynamics and lead to the relaxation rate $\Gamma_{\k}=(\T)^2 D
k^4/(4\eps)$. This internal-mode relaxational rate can be
equivalently rewritten in the form of the center-of-mass
relaxational rate $\Gamma_{\k}=D\T k^2/L(\k)$,
with an effective flux line length $L(\k)={4\eps\over\T k^2}$.
Not surprisingly, $L(\k)$ is approximately the length
in the $z$--direction corresponding to the $xy$--length scale of $k^{-1}$.
The crossover from a dynamics dominated by the center of mass
motion to a dynamics governed by the internal  modes (or equivalently
from $L$ to $L(\k)$) occurs when the wavelength (transverse length scale)
$\sim k^{-1}$ becomes smaller than the {\it static}
transverse ``diffusion'' length
$\sqrt{2\T L/\eps}$ (the 2D projected radius of gyration).
%
%
%
%
%
%

Substituting the expressions for the noninteracting correlation and
response functions
in these two regimes from Appendix C into Eq.\Struct{a}, we obtain the
dynamic interacting structure function in the Gaussian approximation.

\subsec{Center of Mass Dominated Regime}
In the wavevector regime, ${\T \over \eps} \, k^2 L \ll 1$ , where the
center of mass mode dominates we obtain
\eqn\CmStructii{
S(\k,q=0,\omega) \approx {2 n_o D\T k^2 \over
\omega^2 + \left[D\T k^2\left({1\over L} + {n_0 V(\k) \over \T}
\right)\right]^2 }+
{\delta(\w) F(\k,0) n^2_o/(\T)^2\over\left({1\over L} + {n_0 V(\k) \over \T}
\right)^2} \;.}
Concentrating on the first term which describes the dynamics in the absence
of disorder we observe that the center of mass, noninteracting relaxation rate
$\Gamma_{\rm cm} (\k)$  has been additively renormalized by the
interaction between the flux lines. We define an interaction length scale in
the $z-$direction, $L_I(\k)=(n_o V(\k)/\T)^{-1}$, and note that the crossover
between the noninteracting and interacting dynamics occurs when $L>
L_I(\k)$. (see \fig\Linteract{Illustration of the interaction length $L_I(\k)$
defined in the text.})
%
%
%
%

Physically, $L_I(\k)$ is the flux line length (or equivalently
sample thickness) beyond which
there are multiple interactions between the flux lines.
For samples thicker than $L_I(\k)$ (if the crossing barriers are
large) significant line entanglement will take place. This length is therefore
analogous to the entanglement length discussed in Ref.\xref\NS\ .

The expression
for $L_I(\k)$ derived in the Gaussian approximation
will be corrected by the higher order
interactions appearing in the dynamic functional expansion, Eq.\Geff\ .
The flux line dilute regime can be treated using renormalization
group as was first done for the statics in Refs.\xref\NS,\xref\NL.
It is found that in the dilute limit the
interacting theory becomes asymptotically free, with the flux line
interactions renormalizing to zero, logarithmically with length scale.
Following the renormalization group equations down into the dense regime and
matching, gives the effective interaction
$V\approx 4\pi(\T)^2/\eps/\ln(1/n_o\lambda^2)$, which is independent of the
original bare interaction. In the dilute regime (dropping
the unimportant logarithmic factor and constants of order unity) we therefore
obtain $L_I\approx \eps/(2 \T n_o)$. This length is exactly the
entanglement length defined by the static transverse
wandering being on the order of the average line spacing.

For three dimensional samples ($L\rightarrow\infty$), the
flux line density relaxations are dominated by the interactions with the
relaxation rate given by the rate for flux lines with length $L_I(\k)$,
\eqna\RenormDiff$$\eqalignno{\Gamma_{\rm cm}^R&=
{D\T k^2\over L_I(\k)}\;,&\RenormDiff a\cr
&\approx n_o V(\k) D k^2 \;,\qquad\;\;\;\;\; {\rm in\; the\; dense\;
regime}\;,&\RenormDiff b\cr
&\approx {n_o (\T)^2 D k^2\over\eps}\;,\qquad{\rm in\; the\; dilute\;
regime}\;.
&\RenormDiff c\cr}$$

We thus find that while for the noninteracting flux line the center of mass
diffuses very slowly,
with the rate vanishing as $1/L$, the relaxations in the interacting line
liquid are independent of the flux line length $L$.
In the interacting theory the $1/L$ dependence of the diffusion coefficient
gets cutoff by the interaction (entanglement) length $L_I$.
Within the Gaussian approximation
we therefore find that the flux-line
interactions speed up the dynamics of the line liquid, probably because they
lead to a stiffer response to density inhomogeneities.

By comparing with the results of the static structure function
obtained with the Gaussian approximation (see Appendix A and the second part
of this section) we find that flux-line liquid kinetic coefficients are not
modified and the renormalization of the statics is solely
responsible for the increase in the relaxational rate $\Gamma_{\rm cm}$.
Therefore, within the Gaussian approximation no viscosity is generated.
We expect, however, that this will be corrected by higher order interactions,
which also become important in the dense limit, near
$H_{c_2}$. It should be possible to control these interactions via
renormalization group or mode coupling theory.
If the line crossing barriers
remain high, the additional effects of flux line entanglement, not
taken into account by the Gaussian approximation, will play a major role in
the dynamics and probably can
be described by these higher order nonlinearities. We
expect that entanglement and caging effects to become important
in the regime of high line densities and large flux line crossing
barriers, which is possible at intermediate field strengths.
These effects should have a slowing down effect on the dynamics,
and will compete and eventually swamp the relaxational rate
increase found in Eq.\CmStructii\ .

\subsec{Internal Modes Dominated Regime}
In the regime, ${\T \over \eps} \, k^2 L \gg 1$,
\nref\CrossoverComment{In the regime where $L\gg L_I$ the crossover in the
interacting flux-line liquid will occur
at ${\T \over \eps} \, k^2 L_I = 1$.}\CrossoverComment\
where the internal modes dominate we find
\eqnn\Structii$$\eqalignno{
S(\k,q,\w)&=
{ 2 n_o \alpha(\k,q)
  \over
  \w^2 \beta^2(\k,q) + \gamma^2(\k,q)} &\cr
+&\left( {n_o \over \T} \right)^2 \delta(\w) F(k,q)
{ \left(A (\k) b_o (\k,q)
  \right)^2
  \over
  \left[1 + {n_0 V(\k) \over \T}
                 A (\k) b_o (\k,q)
  \right]^2        }
\;,&\Structii\cr}$$
where we have introduced
\eqna\coeffS
$$\eqalignno{
\alpha (\k,q)   &= A(\k) {b_1^2 \over b_3} \Gamma_\k \;,
&\coeffS a \cr
\gamma (\k,q) &= \sqrt{b_1 \over b_3} \Gamma_\k \left[ 1 + {n_o V(\k) \over \T}
                   A(\k) b_0 \right]\;,
&\coeffS b \cr
\beta^2 (\k,q)  &= {b_3 \over b_1} {\gamma^2 (\k,q) \over \Gamma_\k^2}
                     -{b_1 \over b_3 } \,
                   {n_o V(\k) A(\k) \over \T}
                   \biggl[ 2 b_2 + \cr
                   &
                    + {n_o V(\k) A(\k) \over \T}
                   \left( 2 b_0 b_2 - b_1^2 \right) \biggr]\;,
&\coeffS c \cr}$$
and, $A(\k) =  4\epsilon/(\T\sqrt{\pi} k^2)$ and
$\Gamma_\k=D(\T)^2 k^4/(4\eps)$. The coefficients $b_n$ are calculated in
Appendix C and for $2\eps q/(\T k^2)\rightarrow 0$ are constants.
The interacting dynamic structure function above consists of two
parts. The first term is the thermal contribution to the
density-density correlation function and has the standard Lorenzian shape.
We note that the coefficient in front of the $\omega^2$--term depends
nonanalytically on wavevectors $\k$ and $q$ and is therefore a
breakdown of traditional hydrodynamics,
here due to the effects of the internal modes.

The second term in Eq.\Structii\ is the elastic contribution to the dynamic
structure function arising from quenched pinning disorder.
This contribution leads to a time independent persistent contribution to
the structure function. While the density correlations of the unpinned
fraction of lines decay in time as the lines move around, the pinned density
fraction has correlations that are time independent and have spatial
correlations of the random potential.

The linewidth $\gamma (\k)$ has the explicit form
\eqn\gammak{
\gamma(\k) = \sqrt{b_1 \over b_3}
             \left(
             {b_0 \over \sqrt{\pi}} n_0 V(\k) D k^2 + {(\T)^2\over 4 \eps}D k^4
             \right)
}
Note that the $k^4$-term is independent of the interaction between the flux
lines
and is therefore just related to the internal dynamics of a single flux line.
We find that no interaction-generated viscosity appears. However,
flux-line interaction modifies the relaxation rate of a single line
in an important
way. $\gamma(\k)$ describes the crossover of the relaxation rate from the
$(\T)^2 D k^4/(4\eps)$ behavior of a noninteracting line liquid to
$n_0 V(\k) D k^2$ relaxation rate of an interacting liquid. This crossover is
similar to the Bogoliubov crossover in statics \NS.
The interacting dynamics is again diffusive and aside from some constants is
similar to the center of mass dominated regime
considered in beginning of this section.

\subsec{Static Structure Function}
The static structure function $S_s(\k,q)$ is an equal-time
density correlation function, and therefore can be obtained from
the dynamic structure function,
\eqna\StaticRelation$$\eqalignno{S_s(\k,q)&=S(\k,q,t=0)\;,&\StaticRelation a\cr
&=\int {d\w\over 2\pi} S(\k,q,\w)\;.&\StaticRelation b\cr}$$
Applying these equations we obtain the interacting static
structure function of the vortex liquid in the presence of disorder
(see also Appendix A),
\eqna\StaticStruct$$\eqalignno{S_s(\k,q)&={n_o S^o_s(\k,q)\over
1+n_o V(k) S^o_s(\k,q)/\T}
+{F(\k,q)\over (\T)^2}\left({n_o S^o_s(\k,q)\over
1+n_o V(k) S^o_s(\k,q)/\T}\right)^2,\qquad\qquad\;\;&\StaticStruct a\cr
&={n_o\T k^2/\eps\over q^2+(q_B(\k)/\T)^2}+ F(\k,q)
\left({n_o k^2/\eps\over q^2+(q_B(\k)/\T)^2}\right)^2\;,&\StaticStruct b\cr}$$
where $q_B(\k)$ is the Bogoliubov spectrum of the corresponding bosons,\NS
\eqn\Bogoliubov{{q_B(\k)\over\T}=\left[\left({\T k^2\over2\eps}\right)^2+{n_o
V(k) k^2\over\eps}\right]^{1/2}}

The equations for the structure functions that we derived here are valid for
general type of disorder, characterized by disorder correlation function
$F(\k,q)$. The physically relevant cases are:

(i) Point disorder due to oxygen vacancies (uncorrelated)
\eqna\PointDisorder$$\eqalignno{F(\r,z)&=\Delta_0\delta^{(2)}(\r)\delta(z)\;,&
\PointDisorder a\cr
F(\k,q)&=\Delta_0\;.&\PointDisorder b\cr}$$

(ii) Columnar defects (line-correlated along z-axis)
\eqna\LineDisorder$$\eqalignno{F(\r,z)&=\Delta_1\delta^{(2)}(\r)\;,&
\LineDisorder a\cr
F(\k,q)&=\Delta_1\delta(q)\;.&\LineDisorder b\cr}$$

(iii) Grain/twin boundaries (plane-correlated, with
normal $\hat n$ in $xy$-plane)
\eqna\PlaneDisorder$$\eqalignno{F(\r,z)&=\Delta_2\delta(\r\cdot\hat n)\;,&
\PlaneDisorder a\cr
F(\k,q)&=\Delta_2\delta({\hat z}
\cdot(\k\times\hat n))\delta(q)\;.&\PlaneDisorder b\cr}$$

Eq.\Structii\ also depends on the interline interaction $V(k)$
and is valid for a general range of interactions or equivalently for
arbitrary line density.  In the limits of low and high line densities $V(k)$
reduces, respectively, to
\eqna\InteractionLimits$$\eqalignno{V(k)&\approx{\phi_0^2\over8\pi^2}\;,\;\;\;
k\lambda\ll 1\;,&\InteractionLimits a\cr
V(k)&\approx{\phi_0^2\over8\pi^2\lambda^2k^2}=V_o\;,\;\;\;k\lambda\gg 1
\;.&\InteractionLimits b\cr}$$
Eq.\InteractionLimits{a} is valid for $H\approx H_{c1}$ where the lines
are much farther apart than $\lambda$, and
Eq.\InteractionLimits{b} holds for $H>>H_{c1}$ where the average
interline distance is smaller than $\lambda$.

Specializing our general result for the interacting static structure function,
Eq.\StaticStruct\ \ to point disorder and to short-range line interaction,
we recover the result of Nelson and LeDoussal,\NL\
\eqn\NLresult{S_s^{SR}(\k,q)={n_o\T k^2/\eps\over
q^2+(q_B(\k)/\T)^2}+\Delta_0\left({n_o k^2/\eps\over
q^2+(q_B(\k)/\T)^2}\right)^2\;,}
They utilized the boson mapping to obtain this static result. For
pure superconductors
this result reduces to the original result of Nelson and Seung\NS\ also
obtained via the boson mapping.

\newsec{Phenomenological Hydrodynamics}
\seclab\HydroMethod
As explained in the introduction, one can also take a more macroscopic,
phenomenological approach to derive equations of motion for the hydrodynamic
density fields. \MN\  We take this approach in this section with the intent
to subsequently compare the results of our kinetic theory derived
in previous sections with the hydrodynamic approach.

As already emphasized in Sec.\DerivHydro\ , on long time and space scales
the important, slow degrees of freedom that characterize the flux-line liquid
are the number and tangent densities of the vortex liquid,
\eqna\densityA$$\eqalignno{&n(\r,z,t)
=\sum_{i=1}^N\delta^{(2)}\left(\r-\r_i(z,t)\right)\;,&\densityA a\cr
&{\vec t}(\r,z,t)
=\sum_{i=1}^N\pt_z\r_i(z,t)
\delta^{(2)}\left(\r-\r_i(z,t)\right)\;,&\densityA b\cr}$$
together with the current fields $j_a^n(\r,z,t)=n_o v_a(\r,z,t)$,
$j_{a b}^t(\r,z,t)$
\eqna\velocity$$\eqalignno{&j_a^n(\r,z,t)=\sum_{i=1}^N
\pt_t r_{i a}(z,t)\delta^{(2)}\left(\r-\r_i(z,t)\right)
\;,&\velocity a\cr
&j_{a b}^t(\r,z,t)=\sum_{i=1}^N\left[\pt_t r_{i a}(z,t)
\pt_z r_{i b}(z,t)-
\pt_t r_{i b}(z,t)\pt_z r_{i a}(z,t)\right]
\delta^{(2)}\left(\r-\r_i(z,t)\right)\;\qquad\qquad&\velocity b\cr}$$
which transport the number and tangent densities $n(\r,z,t), \, t_a(\r,z,t)$,
respectively. As for the hydrodynamics of a liquid of point particles,
the dynamics of a line-liquid is constrained by dynamic continuity equations
arising from local conservation of the density fields,
\eqna\dContinuity$$\eqalignno{&\pt_t n(\r,z,t)+\pt_a j^n_a(\r,z,t)=0
\;,&\dContinuity a\cr
&\pt_t t_a(\r,z,t)+\pt_b j^t_{a b}(\r,z,t)-\pt_z
j^n_{a}(\r,z,t)=0\;.&\dContinuity b\cr}$$
Furthermore,  $n(\r,z,t)$ and $t_a(\r,z,t)$ are not completely independent.
Continuity of flux lines introduces a spatial constraint between the number
density and the longitudinal part of the tangent density,
\eqn\sContinuity{\pt_z n(\r,z,t)+\pt_a t_a(\r,z,t)=0\;,}
which in the language of bosons plays the role of a temporal continuity
equation for the conservation of the boson density along imaginary
``time'' $z$.
Eqs.\dContinuity\ , \sContinuity\ can be easily verified by substitution using
the microscopic definitions in Eqs.\densityA\ , \velocity . The dynamics
of a flux-line liquid is therefore governed by only two (as opposed to
three) hydrodynamic variables, the number density $n(\r,z,t)$ and
the transverse part of the tangent density ${\vec t}^T(\r,z,t)$.

It is also enlightening to interpret the above continuity equations
in terms of the underlying electromagnetic fields. At length scales much
larger than the London penetration lengths,
$k^{-1} >> \lambda_{\perp}$, $q^{-1} >> \lambda$, Eqs.\Brelation\ \ (or
equivalently, the long wavelength limit of London equation) lead to a simple
relation between the magnetic field and the line densities,
\eqn\BrelationA{{\vec B}(\r,z,t)={\hat z}\phi_o n(\r,z,t) + \phi_o
{\vec t}(\r,z,t)\;.}
It is easy to see that Eqs.\dContinuity\ \ are just the $z$ and $x$-$y$
components of Maxwell equation $\pt_t{\vec B}/c+\nabla\times{\vec E}=0$,
with the electric field ${\vec E}$, which results from
motion of flux lines, related to the
current fields of Eq.\velocity\ .\MN\ The spatial continuity equation,
Eq.\sContinuity\ , is equivalent to ${\vec \nabla}\cdot{\vec B}=0$.

A closed set of hydrodynamic equations is obtained when we supplement the
continuity equations with the constitutive equations for the currents. The
constitutive equations are a statement of Newton's 2nd law, cast in terms of
the hydrodynamic variables. The relations equate the rate of change of the
velocity
to the forces acting on the flux-line liquid. These forces include
the frictional forces, due to flux lines' interaction with the
underlying lattice and weak disorder, and the pressure gradients due to
nonuniformity in the density of the vortex liquid. For simplicity we will
treat the case of small Hall angle and therefore neglect the component of
the velocity response perpendicular to the forces. At long times, and for large
frictional forces the velocity fields quickly decay to their
steady-state value and the inertia term can be ignored. The resulting
time-independent constitutive relation equates the frictional forces,
proportional to velocity fields, to the pressure forces, expressed in terms of
the density fields. A simplified equation for the velocity of the number
density current is given by,
\eqn\constEqn{(\gamma - \eta\nabla^2_{\perp} - \eta_z\pt_z^2) v_a(\r,z,t)=-n_o
\pt_a{\delta\H\over\delta n(\r,z,t)}+n_o\pt_z{\delta\H\over\delta
t_a(\r,z,t)}+f^{ext}_a(\r,z,t)+{\nu}_a(\r,z,t)\;.}
The terms on the left hand side of the above equation represent the frictional
and viscous forces acting on the flux lines, which are balanced by the pressure
gradient forces, the external forces ${\vec f}_{ext}$ and a random noise force
appearing on the right hand side of the equation.
${\vec f}_{ext}$  might include the Lorentz force,
${\vec f}_{ext}=-n_o\phi_o{\hat z}\times {\vec j_e}$ due to the charge
current ${\vec j_e}$ coupling to the magnetic field of the flux-line liquid.
Here we are interested in the equilibrium regime, ${\vec f}_{ext}=0$.
We take ${\nu}_a(\r,z,t)$ to be a Gaussian zero-mean noise with
the correlations determined by the Fluctuation Dissipation Theorem,
\eqna\NoiseEffnu$$\eqalignno{\langle{\nu}_a(\k,q,\w){\nu}_b
(\k',q',\w')\rangle&=&\NoiseEffnu a\cr
= \T\left(\gamma + \eta\k^2 +
\eta_z q^2\right)&(2\pi)^4\delta^{(2)}(\k+\k')
\delta(q+q')\delta(\w+\w')\delta_{a b}\;,\cr
\langle{\nu}_a(\k,q,\w)\rangle&=0\;.&\NoiseEffnu b\cr}$$

For the hydrodynamic description it is sufficient to assume that the
effective Hamiltonian, $\H[n,{\vec t}]$ is an expansion in powers of
hydrodynamic variables $n(\r,z,t)$ and ${\vec t}(\r,z,t)$, which
for simplicity we truncate at the quadratic order. It is more convenient to
expand in terms of the fluctuations $\rho(\r,z)=n(\r,z)-n_o$ and
$\Ttau(\r,z)={\vec t}(\r,z)-{\vec t}_o$ around the average values $n_o$,
${\vec t}_o$, so that the linear terms can be eliminated.
Ignoring disorder (because it effects can be easily included)
and assuming translational invariance, we obtain in Fourier
space,
\eqn\Hhydro{\H[\rho(\k,q),\Ttau(\k,q)]=\int_{\k,q}{1\over2}
\left[K_1(\k,q)\rho^2+K_2(\k,q) {\Ttau}^2\right]\;,}
where functions $K_1(\k,q)$ and $K_2(\k,q)$ are related to the static
density correlation functions. Imposing the constraint of Eq.\sContinuity\
in Fourier space, $q\rho(\k,q)=-\k\cdot\Ttau(\k,q)$ to reexpress
$\H$ in terms of independent hydrodynamic variables, we easily compute
the static structure function, $\langle\rho(\k,q)\rho(-\k,-q)\rangle$, and
the tangent correlation function $\langle\tau_a(\k,q)\tau_b(-\k,-q)\rangle$.
\eqna\HydroCorrFunc$$\eqalignno{S^s(\k,q)&={\T k^2\over k^2 K_1(\k,q)+q^2
K_2(\k,q)}\;,&\HydroCorrFunc a\cr
T^s_{a b}(\k,q)&=P^T_{a b}(\k){\T\over K_2(\k,q)}+
P^L_{a b}(\k){q^2\over k^2}S^s(\k,q)\;.&\HydroCorrFunc b\cr}$$
This allows us to reexpress Eq.\Hhydro\ in terms of the static
structure function $S^s(\k,q)$ and the transverse part of the tangent
density correlation function $T_T(\k,q)$,
\eqn\HhydroST{\H[\rho(\k,q),\Ttau(\k,q)]=\int_{\k,q}{\T\over2}
\left[\left(S_s^{-1}(\k,q)-{q^2\over k^2} T_T^{-1}(\k,q)\right)\rho^2+
T_T^{-1}(\k,q){\Ttau}^2\right]\;.}

Combining the above equation with Eqs.\constEqn\ , \sContinuity\ ,
\dContinuity{a}\ we find that the dynamic equations for $n(\k,q,t)$ and
${\vec t}(\k,q,t)$ decouple. Here we will only concentrate on the
hydrodynamics of $n(\k,q,t)$ which is governed by a simple Langevin equation,
\eqn\HydroDynEqn{\pt_t n + \left(n_o^2\T D(\k,q) k^2 S_s^{-1}(\k,q)\right) n =
{\hat\zeta}(\k,q,t)\;,}
where we defined the hydrodynamic diffusion coefficient
$D^{-1}(\k,q)=\gamma + \eta \k^2 + \eta_z q^2$ and the noise
${\hat\zeta}(\k,q,t)=i n_o D(\k,q)\k\cdot{\vec{\nu}}(\k,q)$,
which is also Gaussian with zero-mean and the correlations determined by those
of
$\vec{\nu}$, Eq.\NoiseEffnu\ ,
%
\eqna\NoiseEffzeta$$\eqalignno{\langle{\hat\zeta}(\k,q,\w){\hat\zeta}(\k',q',\w')\rangle
&=\T D(\k,q) k^2 n_o^2 (2\pi)^4\delta^{(2)}(\k+\k')
\delta(q+q')\delta(\w+\w')\;,\qquad\qquad\qquad&\NoiseEffzeta a\cr
\langle{{\hat\zeta}}(\k,q,\w)\rangle&=0\;.&\NoiseEffzeta b\cr}$$
These equations then lead to the hydrodynamic structure and response
functions previously derived in Ref.\xref\MN.
\eqna\CorrFuncPhenom
$$\eqalignno{
S^{phenom} (\k,q,\omega)
&={2 n^2_o\T k^2 D(\k,q)\over \omega^2 + \left(n_o^2\T k^2
D(\k,q)/S_s(\k,q)\right)^2} \;,&\CorrFuncPhenom a\cr
\S^{phenom} (\k,q,\omega)
&={n^2_o\T k^2 D(\k,q)\over i\omega + n_o^2\T k^2 D(\k,q)/S_s(\k,q)}
\;,&\CorrFuncPhenom b\cr}$$

We are now in a position to compare the results of the phenomenological
model of hydrodynamics with our kinetic theory of the flux-line liquid.
Making the comparison in the physically most relevant regime dominated by
center-of-mass motion, Eq.\CmStructii\ , we find the following
identifications,
\eqna\Identify$$\eqalignno{
D &\longleftrightarrow n_o D(\k,q=0)\;,&\Identify a\cr
D\left({1\over L}+{n_o V(\k)\over\T}\right)&\longleftrightarrow
{n_o D(\k,q=0)\over S_s(\k,q=0)}\;.&\Identify b\cr}$$
As we already observed in Sec.\Details\  no viscosity is generated within
the Gaussian approximation employed here and we are only able to establish a
simple relations between the parameters of our kinetic model and the
phenomenological theory,
\eqna\Identifyii$$\eqalignno{
\gamma &\longleftrightarrow {n_o\over D}\;,&\Identifyii a\cr
S_s &\longleftrightarrow n_0 \left({1\over L}+{1\over L_I(\k)}\right)^{-1}
\;.&\Identifyii b\cr}$$
Eq.\Identifyii{a} is physically appealing in its identification of
the phenomenological friction coefficient $\gamma$ with the inverse
of the kinetic diffusion coefficient $D$. Eq.\Identifyii{b} leads to the
the static structure function, in agreement with the result obtained from boson
analogy.
The above equations show that at least within the Gaussian approximation,
in the center-of-mass dominated regime the dynamics is modified only through
the statics, i.e. the kinetic coefficients are not renormalized in this order
of approximation.
\newsec{Conclusions}
In this paper we have formulated
the dynamic theory of the flux line liquid phase directly from the kinetic
theory
of individual, interacting flux lines. We have used the resulting theory to
study
the dynamic structure and response functions of the line liquid in the
presence of various types of pinning disorder. In order to solve the theory we
employed a Gaussian approximation to the dynamic functional, which should be
valid at intermediate flux line densities or fields $H_{c_1}\ll H\ll H_{c_2}$,
where the effects of large fluctuations are not as important.

We expressed the interacting dynamic structure function in terms of the single
line structure functions. In the long time limit, $t\gg
t_{Rouse}$ and/or for transverse wavelengths larger than the line wandering,
the center of mass mode dominates over the internal mode, and we
recover the hydrodynamics of rigid rods. While for noninteracting
lines the diffusion coefficient
vanishes as $L\rightarrow\infty$, the interactions between the lines
lead to an increase in the relaxation rate of the line liquid. The
diffusion rate remains finite for any $L$, with $L$ cutoff by the interaction
length $L_I(\k)$. In the dense limit $L_I$ is determined by the interactions,
with the two-body interaction giving only a first order estimate to
this length. A detailed calculation that takes into account higher order
interactions is required in this dense limit. In the dilute regime, the
renormalization
group calculations lead to a line interaction that vanishes as an
inverse of a logarithm of the length scale and the microscopic interaction
drops out. In this case $L_I$
becomes just the entanglement length defined in Ref.\xref\NS .
In either case for, $t\gg t_{Rouse}$,
our results are therefore in agreement with the
phenomenological model of Marchetti and Nelson, but unfortunately the Gaussian
approximation is not accurate enough to generate
the viscosity of their model from our kinetic theory.

In the opposite limit of short times and/or ${\T k^2 L\over\eps}\gg 1$,
we find that the
internal modes dominate. In the presence of interactions we find that
the $k^4$ relaxation rate of a noninteracting flexible line liquid crosses over
to a
diffusive $k^2$ relaxation rate. In contrast with the
phenomenological model, however, a
wavevector-dependent coefficient of the $\w^2$ term is generated. This
coefficient $\sim k^{-4}$ and by this
constitutes a break down of conventional hydrodynamics due to the internal
mode fluctuations.

We find that the Gaussian approximation successfully reproduces the static
Bogoliubov ``spectrum'' previously obtained via boson analogy. For the
statics this approximation is therefore equivalent to an infinite summation
of all ladder graphs in the single line picture, and then treating the
resulting field theory in the mean-field approximation. The dynamic results of
this approximation have the structure of a dynamic Random Phase Approximation
(RPA). In addition to the effects discussed above this
approximation leads to a disorder generated
perturbative correction to the dynamic structure
function, which agrees with the phenomenological
approach.

It is well known that preserving the discreteness of the flux lines is crucial
for
the correct treatment of the low temperature ordered phases like the vortex
glass and Bose glass. It is likely that in the liquid phase,
however, much of the dynamics can be described in terms of the density fields
which coarse-grain over the discrete line coordinates as we have done here.
In this description slow dynamics for high line crossing barriers
and entanglement can in principle be incorporated by higher order nonlocal
interactions in the densities, although only in some average sense. Also
line crossing and recombination are present in the theory and can be
controlled in an average sense by the strength of the repulsive flux-line
interaction. This description of a flux-line liquid can be improved by
also introducing a local tangent field that keeps track of the
local orientation of the flux lines as we do for the statics in Appendix A.

We expect that the results derived in this paper will be corrected by the
higher order interaction which can be taken into account using
renormalization group or mode coupling methods. These
corrections will undoubtly lead to renormalization of the
kinetic coefficients. We also expect that the slow dynamics resulting
from the lines being caged by its neighbors and from flux line entanglement
effects,
can be reproduced by these higher order interactions. It is quite likely,
however,
that these effects will turn out to be non-perturbative.
It is also possible that the $k^2$ correction to the diffusion coefficient
will be absent even beyond the Gaussian approximation. The interactions might
generate only nonanalytic terms in $k$ and therefore the flux-line liquid
viscosity will be strictly zero, leading to a breakdown of the
phenomenological model described in Sec.8.
In any case we believe that the formulation of hydrodynamics in
terms of kinetic theory of
lines introduced in this paper will be useful for a more detailed
understanding of flux-line liquid phase at and possibly away from equilibrium.
\vfill\eject
\noindent
{\bf Acknowledgements}

We would like to thank professors David Nelson, Daniel Fisher and
Bert Halperin for numerous stimulating discussions and for the critical
reading of the manuscript. This work was supported by National
Science Foundation, through Grant No. DMR91--15491 and through the
Harvard Materials
Research Laboratory. Leo Radzihovsky acknowledges support from
Hertz graduate fellowship. The work of Erwin Frey has been supported by
the Deutsche Forschungsgemeinschaft (DFG) under Contracts No. Fr. 850/2-1,2.
\vfill\eject
\appendix{A}{Calculation of the Interacting Static Structure Function}
In this Appendix we derive the interacting static structure function of
Eq.\StaticStruct\ \ within the static equilibrium formulation, directly
from the
microscopic Hamiltonian Eq.\Hamiltonian, using methods similar to the ones
used in the main text to obtain the dynamic structure function.

We begin with the microscopic Hamiltonian,
\eqn\Hamiltonianii{\H={\epsilon\over 2}\sum_{i=1}^{N}\int_0^L
dz\left(\pt_z \r_i\right)^2+{1\over 2}\sum_{i\neq j=1}^N\int_0^L
dz V(\r_i(z)-\r_j(z))+\sum_{i=1}^N\int_0^L dz U(\r_i(z),z)\;.}
Analogously to the dynamics calculation we want to derive a macroscopic
Hamiltonian in terms of number and tangent densities,
$\rho(\r,z)=n(\r,z)-n_0$ and $\vec\tau(\r,z)={\vec t}(\r,z)-{\vec t}_0$, where,
\eqna\densityii$$\eqalignno{&n(\r,z)
=\sum_{i=1}^N\delta^{(2)}\left(\r-\r_i(z)\right)\;,&\densityii a\cr
&{\vec t}(\r,z)
=\sum_{i=1}^N{\pt\r_i\over\pt z}
\delta^{(2)}\left(\r-\r_i(z)\right)\;,&\densityii b\cr}$$
The tangent density field can be written as a sum of its longitudinal and
transverse parts, $\vec{\tau}(\r,z)=\vec{\tau}^L(\r,z)+\vec{\tau}^T(\r,z)$
with $\tau^L_a(\r,z)=P^L_{a b}\tau_b(\r,z)$ and
$\tau^T_a(\r,z)=P^T_{a b}\tau_b(\r,z)$,
where $P^T_{a b}(\k)=\delta_{a b}-k_a k_b/k^2$
and $P^L_{a b}(\k)=k_a k_b/k^2$, in Fourier space.

As already mentioned in the derivation of dynamics, the longitudinal part of
the tangent density field $\vec{\tau}^L(\r,z)$ is related to the number
density $n(\r,z)$ by the ``continuity equation'',
\eqn\Continuity{{\pt\over\pt z}\rho(\r,z)+
\vec{\nabla}_r\cdot\vec{\tau}(\r,z)=0\;,}
which is a statement that the flux lines do not end within the sample, or
equivalently in terms of the magnetic field is Maxwell equation,
$\vec{\nabla}\cdot\vec{B}(\r,z)=0$. This equation is clearly trivial for
$\vec{\tau}^T(\r,z)$, but gives a relation between $\rho(\r,z)$ and
$\vec{\tau}^L(\r,z)$.  Clearly then it is not necessary to keep track of the
longitudinal part of the tangent density since its correlations can be easily
obtained from those of the number density using Eq.\Continuity\ .
We derive an effective Hamiltonian in terms of static density fields
$\rho(\r,z)$ and $\vec{\tau}^T(\r,z)$
by starting with the microscopic Hamiltonian
Eq.\Hamiltonianii\ and integrating out microscopic degrees of freedom
$\r_i(z)$ as we did for the dynamic calculation.

The microscopic Hamiltonian can be rewritten (aside from irrelevant
constants) in the suggestive form in terms
of the relevant density fields,
\eqn\Hamiltonianiii{\H[\r_i]=\H_o[\r_i]+\H_{int}[\rho,\vec{\tau}]\;,}
where $\H_o[\r_i]$ is the Hamiltonian of $N$ noninteracting
lines, the first term in Eq.\Hamiltonianii\ and,
\eqn\DensityHamiltonian{\H_{int}[\rho,\vec{\tau}]=
\int_{\r,\r\ ',z}\rho(\r,z)\rho(\r\ ',z) V(\r-\r\ ')+
\int_{\r,z}\rho(\r,z)U(\r,z)\;.}
Since we are looking for a description in terms of the density fields, we
construct a generating functional by introducing external fields $h(\r,z)$ and
$\h_\t(\r,z)$ that couple to $\rho(\r,z)$ and $\vec{\tau}^T(\r,z)$,
respectively,
\eqn\Gfuni{Z_d[h,\h_\t]=\int{\D}\r_i(z)
\exp\left[-{1\over\T}\H[\r_i(z)]+
\int_{\r,z}\left(\rho(\r,z)
h(\r,z)+\vec{\tau}(\r,z)\cdot\h_\t(\r,z)\right)\right]}

We introduce unity inside above expression in the form of functional
$\delta$-functions, constraining the auxiliary fields $\psi(\r,z)$ and
$\Ppsi_\t(\r,z)$ to equal the physical densities fluctuations, respectively,
\eqnn\HamiltonianEff
$$\eqalignno{Z_d[h,\h_\t]
&=\int{\cal D}\r_i(z)
e^{-\beta\H_o[\r]-\beta\H_{int}[\rho,\vec{\tau}]}
e^{\int_{\r,z}\left(\rho(\r,z) h(\r,z)+\vec{\tau}(\r,z)\cdot\h_\t(\r,z)\right)}
&\cr
&\times
\int{\cal D}\psi{\cal
D}\Ppsi_\t\prod_{\r,z}\delta\left(\psi(\r,z)-\rho(\r,z)\right)
\delta\left(\Ppsi_\t(\r,z)-\vec{\tau}^T(\r,z)\right)&\cr
& &\cr
&=\int{\cal D}\psi{\cal D}\Ppsi_\t
e^{-\beta\H_{int}[\psi,\Ppsi_\tau]}
e^{\int \psi h + \int \Ppsi_\tau \cdot \h_\t} &\cr
&\times
\left\langle
\delta\left(\psi(\r,z)-\rho(\r,z)\right)
\delta\left(\Ppsi_\t(\r,z)-\vec{\tau}^T(\r,z)\right)\right\rangle_o\;,
&\HamiltonianEff\cr}$$

The interactions and disorder are quadratic and linear functions of auxiliary
density fields and the nontrivial part of the calculation reduces to averaging
the functional $\delta$-functions with the Boltzmann weight $\exp[-\beta\H_o]$
of $N$ noninteracting lines. These averages decouple to single line averages,
\eqnn\delfuncii$$\eqalignno{&\left\langle\delta\left(\psi(\r,z)
-\rho(\r,z)\right)
\delta(\Ppsi_\t(\r,z)-\vec{\tau}^T(\r,z))\right\rangle_o
=\cr
&\;\;\;\;\;\;\;\;\;=\int{\cal D}\phi{\cal D}\Pphi_\t
e^{i\int(\phi\psi+\Pphi_\t\cdot\Ppsi_\t)}\left[\left\langle
e^{-i\int(\phi\rho_1+
\Pphi_\t\cdot\vec{\tau}_{t1}}\right\rangle_o\right]^N\;.&\delfuncii\cr}$$

Inserting above equation into Eq.\HamiltonianEff\ \ we obtain the partition
function (for a fixed realization of disorder) in terms of an effective
Hamiltonian expressed as an expansion in macroscopic density fields,
\eqn\Ziii{Z_d[h,\h_\t]=
\int{\cal D}\psi{\cal D}\Ppsi_\t
\int{\cal D}\phi{\cal D}\Pphi_\t
\exp\left[-\H[\psi,\Ppsi_\t,\phi,\Pphi_\t]+\int(\psi h+
\Ppsi_\t\cdot\h_\t)\right]\; ,}
where the effective Hamiltonian is given by,
\eqn\Heff{\H[\psi,\Ppsi_\t,\phi,\Pphi_\t]=\H_{int}[\psi]+
n_o\Gamma[\phi,\Pphi_\t]+i\int(\phi\psi+\Pphi_\t\cdot\Ppsi_\t)\;,}
with the new contribution $n_o\Gamma[\phi,\Pphi_\t]$
coming from the average of the
$\delta$-functions in Eq.\delfuncii\ and is expressed in terms of a single
flux line cumulant expansion of $\rho_1$ and $\vec{\tau}_1^T$, in Fourier
space,
\eqnn\Geffii $$\eqalignno{&-\Gamma[\phi,\Pphi_\t]=
\sum_{l,m}^\infty{(-i)^{(l+m)}\over l! m!}
\int_{q_1,\k_1}\cdots\int_{q_{l+m},\k_{l+m}}
\Gamma^{(l,m)}_{a_1\cdots a_m}(\k_1,q_1\cdots\k_{l+m},q_{l+m})
\times&\cr
&\times(2\pi)^3\delta^{(2)}\left(\sum_{i=1}^{l+m}\k_i\right)
\delta\left(\sum_{i=1}^{l+m}q_i\right)\times &\cr
&\times\phi(-\k_1,-q_1)\cdots\phi(-\k_l,-q_l)
\phi_{\t a_1}(-\k_{l+1},-q_{l+1})
\cdots\phi_{\t a_m}(-\k_{l+m},-q_{l+m})\;,&\cr
&&\Geffii\cr}$$
where the sum over $a_i$ is implied.
$\Gamma^{(l,m)}_{a_1\cdots a_m}$
is a connected correlation function of a single line density fluctuations
$\rho_1$ and $\vec{\tau}_1^T$, and in Fourier space is,
\eqnn\Gammasii$$\eqalignno{\Gamma^{(l,m)}_{a_1\cdots a_m}
&(\k_1,q_1\cdots\k_{l+m},q_{l+m})\;(2\pi)^3\delta^{(2)}
\left(\sum_{i=1}^{l+m}\k_i\right)\delta\left(\sum_{i=1}^{l+m}q_i\right)&\cr
&=A\langle\rho(\k_1,q_1)\cdots\rho(\k_l,q_l)
\tau^T_{a_1}(\k_{l+1},q_{l+1})\cdots\tau^T_{a_m}(\k_{l+m},q_{l+m})
\rangle_o\;.\qquad\qquad&\Gammasii\cr}$$
As in dynamics we have inserted an area factor $A$ for convenience and
dropped the single line label $1$.
The vertex functions $\Gamma^{(l,m)}_{a_1\cdots a_m}$
can be
systematically computed for any $l,m$. We calculate some of the vertex
functions for both statics and dynamics
in Appendices B and C.

Within the Gaussian approximation we truncate $\Gamma[\phi,\Pphi_\t]$
at quadratic order in the densities. We observe that
$\Gamma^{(2,0)}(\k,q)=S^o_s(\k,q)$,
$\Gamma^{(0,2)}_{a b}(\k,q)=T^o_{s ab}(\k,q)$
and $\Gamma^{(1,1)}_a(\k,q)$ are single line number and tangent
static structure functions (with notation $S^o_s(\k,q)$ and
$T^o_{s ab}(\k,q)$ from Ref.\NS\ ),
derived in Appendix B and discussed in the main text,
\eqna\noninterStaticii$$\eqalignno{\Gamma^{(2,0)}(\k,q)
&={k^2\T/\eps\over q^2+(k^2\T/2\eps)^2}\;,&\noninterStaticii a\cr
\Gamma^{(0,2)}_{a b}(\k,q)
&={\T\over\eps}P^T_{a b}\;,&\noninterStaticii b\cr
\Gamma^{(1,1)}_a(\k,q)
&=0\;.&\noninterStaticii c\cr}$$
The fact that $\Gamma^{(1,1)}_a(\k,q)$ and all the cross correlation functions
vanish for the transverse part of the tangent density field $\tau^T_a$ is
a result of the independence of $\tau^T_a$ from $\rho$. Mathematically
it is true to all orders because any cross correlation function of $\tau_a$
is a tensor
with atleast one index $a$ which must be carried by $k_a$. This means
that all the cross correlation functions are purely longitudinal and when
contracted with $P^T_{a b}$ (for every $\tau^T_a$) automatically vanish.

Integrating over the auxiliary fields $\phi(\k,q)$ and $\Pphi_\t(\k,q)$ we
obtain an effective Hamiltonian expressed solely in terms of the physical
number and transverse tangent density fields,
\eqnn\Heffii$$\eqalignno{\H[\psi,\Ppsi]&=
\int_{\k,q}\left[{1\over2}\psi(\k,q)A^{-1}(\k,q)\psi(-\k,-q)+
{1\over2}\psi^{\tau}_a(\k,q)B^{-1}_{a b}(\k,q)\psi^{\tau}_b(-\k,-q)\right. &\cr
&\;\;\;\;\;\;\;\;\;\;\;\;\;\;\;\;\;\;\;\;
\left.+\psi(\k,q) U(-\k,-q)\right]\;,&\Heffii\cr}$$
where $A(\k,q)$ and $B_{a b}(\k,q)$ are the static interacting but
disorder-free 2pt-correlation functions of $\rho$ and $\tau^T_a$,
\eqna\StaticStructApp$$\eqalignno{A(\k,q)&=
{n_o\T k^2/\eps\over q^2+(q_B(\k)/\T)^2}\;,&\StaticStructApp a\cr
B_{a b}(\k,q)&=P^T_{a b} {\T n_o\over\eps}\;,&\StaticStructApp b\cr}$$
and $q_B(\k)$ is the Bogoliubov spectrum given by Eq.\Bogoliubov\  in
the main text.

We use above Hamiltonian to
derive the interacting static functions in the presence of disorder,
$S^s(\k,q)=\overline{\langle\rho(\k,q)\rho(-\k,-q)\rangle}_o$ and
$T^s_{a b}(\k,q)=\overline{\langle\tau_a(\k,q)\tau_b(-\k,-q)\rangle}_o=
T_s^T(\k,q) P^T_{a b}+T_s^L(\k,q) P^L_{a b}$.
The longitudinal part of $T^s_{a b}$
can be obtained from $S^s(\k,q)$ by using flux line continuity
Eq.\Continuity\ . Averaging over the annealed density fields, followed by
the average over the quenched disorder we obtain,
\eqna\StaticStructAppii$$\eqalignno{S_s(\k,q)&=
{n_o\T k^2/\eps\over q^2+(q_B(\k)/\T)^2}+ F(\k,q)
\left({n_o k^2/\eps\over
q^2+(q_B(\k)/\T)^2}\right)^2\;,\qquad\qquad&\StaticStructAppii a\cr
T_s^T(\k,q)&={\T n_o\over\eps}\;,&\StaticStructAppii b\cr
T_s^L(\k,q)&={q^2\over k^2} S_s(\k,q)\;.&\StaticStructAppii c\cr}$$
Eq.\StaticStructAppii{a} is in complete agreement with static limit
of the interacting dynamic structure function obtained in the main
text, Eq.\StaticStruct{b}. For short-range disorder and flux
line interaction, $F(\k,q)=\Delta_0$ and $V(k)=\phi_0^2/8\pi^2$,
respectively above equations reduce to the results of Ref.\xref\NL\
obtained using the boson mapping.
\vfill\eject
\appendix{B}{Calculation of the Nonlinearities in the Static
Cumulant Expansion}
In this appendix we calculate the coefficients of the nonlinear interactions
which appear in the cumulant expansion of the Hamiltonian in Appendix A.

We begin with static problem and calculate the static single line correlation
functions of the density fields $\rho$ and $\tau^T\a$,
defined by Eq.\Gammasii\ . It is convenient to work in the space of $(\k,z)$.
We evaluate the following static correlation function,
\eqna\GammasApp$$\eqalignno{\Gamma^{(l,m)}_{a_1\cdots a_m}
&(\k_1,z_1\cdots\k_{l+m},z_{l+m})\;{(2\pi)^2\over A}\delta^{(2)}
\left(\sum_{i=1}^{l+m}\k_i\right)&\cr
&=\langle\tau^T_{a_1}(\k_1,z_1)\cdots\tau^T_{a_m}(\k_m,z_m)
\rho(\k_{m+1},z_{m+1})\cdots\rho(\k_{m+l},z_{m+l})
\rangle_o\;,\qquad\qquad&\GammasApp a\cr
&=\langle\pt_{z_1} r_{a_1}(z_1)\cdots\pt_{z_m} r_{a_m}(z_m) e^{-i(
\k_1\cdot\r(z_1)+\ldots+\k_{l+m}\cdot\r(z_{l+m})}\rangle_o^T\;.
&\GammasApp b\cr}$$
where the averages are performed with a Boltzmann weight of a single line and
the superscript $T$ extracts the transverse part of the average.

Above average can be easily computed by using the following Lemma
that holds for Gaussian averages,
\eqn\Lemma{\langle X_1\cdots X_l e^Y\rangle_o =\left[\sum_{i\neq j}^l
\langle X_i X_j\rangle_o+\langle Y X_1\rangle_o\cdots
\langle Y X_l\rangle_o\right] e^{\langle Y^2\rangle_o/2}}
where, $X_i$ and $Y$ are Gaussian random variables with respect to the
measure used. Above equation can be easily proved by rewriting the
left-hand-side as an multiderivative with respect to $l$ parameters
of an exponential generating function,
\eqna\Proof$$\eqalignno{\langle X_1\cdots X_l e^Y\rangle_o
&=\pt_{\mu_1}\cdots\pt_{\mu_l}\left|_{\{\mu_i\}=0}\right.
\langle e^{Y+\mu_i
X_i}\rangle_o\;,&\Proof a\cr
&=\pt_{\mu_1}\cdots\pt_{\mu_l}\left|_{\{\mu_i\}=0}\right.
e^{\langle Y^2\rangle_o/2+\langle Y\mu_i
X_i\rangle_o+\langle(\mu_i X_i)^2\rangle_o/2}\;,\qquad\qquad&\Proof b\cr}$$
where we used a property of averages over Gaussian variables,
$\langle\exp(\phi)\rangle_o=\exp(\langle\phi^2\rangle_o/2)$. Performing the
differentiation and evaluating the result at $\{\mu_i\}=0$ we obtain
Eq.\Lemma\ .

Returning to the original problem, Eq.\GammasApp\ \ , we make the
identification
$X_i=\pt_{z_i} r_a(z_i)$ and $Y=-i\sum_{i=1}^{l+m}\k_i\cdot\r(z_i)$,
obtaining,
\eqna\YY$$\eqalignno{\langle Y^2\rangle_o&=-\left\langle
\left(\sum_{i=1}^{l+m}\k_i\cdot\r(z_i)\right)^2\right\rangle_o\;,&\YY a\cr
&=-{1\over 2}\sum_{i=1}^{l+m}k^2_i\langle r^2(z_i)\rangle_o-
\sum_{i>j}^{l+m}\k_i\cdot\k_j\langle \r(z_i)\cdot\r(z_j)\rangle_o\;,&\YY b\cr
&=-\left(\sum_{i=1}^{l+m} k_i\right)^2 C(0)-
2\sum_{i>j}^{l+m}\k_i\cdot\k_j\left[C(z_i-z_j)-C(0)\right]\;,&\YY c\cr}$$
where the $\r(z_i)$ averages are isotropic,
\eqna\rAverages$$\eqalignno{\half\langle
\r(z_i)\cdot\r(z_j)\rangle_o&=C(z_i-z_j)\;,&\rAverages a\cr
&=\int_q {e^{i q (z_i-z_j)}\over\eps q^2}\;.&\rAverages b\cr}$$
Using this expression we find,
\eqna\rAveragesii$$\eqalignno{C(z_i-z_j)-C(0)&=
\int_q {e^{i q (z_i-z_j)}-1\over\eps q^2}\;,&\rAveragesii a\cr
&=-{|z_i-z_j|\over 2\eps}\;.&\rAveragesii
b\cr}$$
 From Eq.\rAverages{b}\ we observe that $C(0)\approx L\approx
A\rightarrow\infty$ and therefore this first factor leads to a
$\delta$-function as $L, A\rightarrow\infty$ and imposes momentum
conservation in the $xy$-plane for all the correlation functions.

The $\langle X_i Y\rangle$ and $\langle X_i X_j\rangle$ averages
can be similarly evaluated,
\eqna\XY$$\eqalignno{\langle X_i Y\rangle_o&=
-i\pt_{z_i}\sum_{j=1}^{l+m}\langle\r(z_i)\k_j\cdot\r(z_j)\rangle_o\;,&\XY a\cr
&=-i\sum_{j=1}^{l+m}\k_j g(z_i-z_j)\;,&\XY b\cr}$$
where,
\eqna\gFunction$$\eqalignno{g(z_i-z_j)&=\half\pt_{z_i}\langle\r(z_i)
\cdot\r(z_j)\rangle_o\;,&\gFunction a\cr
&=-{2\over\eps}\int_0^\infty{d q\over 2\pi}{\sin q(z_i-z_j)\over
q}\;,&\gFunction b\cr
&=-{1\over 2\eps}sgn(z_i-z_j)\;,&\gFunction c\cr}$$
with $sgn(z)=1$ for $z>0$ and $sgn(z)=-1$ for $z<0$.
\eqna\XX$$\eqalignno{\langle X_i X_j\rangle_o&=
\pt_{z_i}\pt_{z_j}\langle\r(z_i)\cdot\r(z_j)\rangle_o\;,&\XX a\cr
&=2\int_q e^{i q(z_i-z_j)}\;,&\XX b\cr
&=2\delta(z_i-z_j)\;.&\XX c\cr}$$

Combining above equations inside Eq.\GammasApp\ \ we find for odd $m$
the cumulants vanish and for even $m$ we obtain,
\eqnn\GammasAppii$$\eqalignno{\Gamma^{(l,m)}_{a_1\cdots a_m}&
(\k_1,z_1\cdots\k_{l+m},z_{l+m})&\cr
&=\left({1\over\eps}\right)^{m/2}\left[P^T_
{a_1 a_2}\delta(z_1-z_2)P^T_{a_3 a_4}\delta(z_3-z_4)\cdots
P^T_{a_{m/2-1} a_m/2}\delta(z_{m/2-1}-z_{m/2})\right.&\cr
&\left.+{\rm All\;pair\;combinations}\right]
\times\exp\left[{\T\over2\eps}\sum_{i>j}^{l+m}\k_i\cdot\k_j
|z_i-z_j|\right]\;.\qquad&\GammasAppii\cr}$$
For Gaussian approximation that we are concerned with in the main text we
obtain,
\eqna\GammaTwo$$\eqalignno{\Gamma^{(2,0)}
(\k,z_1,-\k,z_2)&=e^{-{\T\over2\eps}k^2 |z_1-z_2|}\;,&\GammaTwo a\cr
\Gamma^{(0,2)}_{a_1 a_2}(\k,z_1,-\k,z_2)&=
{\T\over\eps}P^T_{a_1 a_2}\delta(z_1-z_2)
e^{-{\T\over2\eps}k^2 |z_1-z_2|}\;,&\GammaTwo b\cr
\Gamma^{(1,1)}_{a_1}(\k,z_1,-\k,z_2)&=0\;.&\GammaTwo c\cr}$$
These expressions when Fourier transformed to $q$ space give
Eqs.\noninterStaticii\ \ of Appendix A.
\vfill\eject
\appendix{C}{Noninteracting Flux-Line Liquid Dynamics}
In this appendix we apply methods similar to Appendix B to calculate the
dynamic cumulants
\eqn\GammasAppC{
\Gamma^{(2)}_{a b} (x_1,x_2)
=\langle \rho_{a}(x_1) \rho_{b}(x_2) \rangle_o \;,}
where  $x=(\k,z,t)$ and $a,\, b=1,2$.
The procedure is similar to the one used for statics.
The averages are  performed with the exponential of
the single flux line dynamic functional $\J_o[\rr,\r]$, Eqs.\Jo-\RRG\ ,
instead of with the static Boltzmann factor. As explained in the main text
in the derivation of hydrodynamics we traced over the tangent density field
and its corresponding response field. The longitudinal part of the dynamic
tangent correlation function can be extracted from those of the number density
field by using flux line continuity Eq.\Continuity\ , while as for the
statics, the transverse part of the tangent correlation function is probably
not
affected by the flux line interaction or disorder.

In order to calculate the averages in Eq. \GammasAppC\ we introduce ``Rouse
modes'',
\eqna\RouseAppC
$$\eqalignno{
\r(q,t) &= {1 \over L} \int_0^L dz \cos (p_q z) \r(z,t) \;,&\RouseAppC a\cr
\r(z,t) &=\r_o (t) + 2 \sum_{q \geq 1} \cos (p_q z) \r(q,t) \;,&\RouseAppC b\cr
}$$
where $p_q = {\pi q \over L}$ and $L$ is the length of the flux line.
$\r_o (t)$ denotes the center of mass mode.
 From the single line dynamic functional $\J_o[\rr,\r]$, Eqs.\Jo-\RRG\ ,
one can derive the following correlation functions for the ``Rouse modes'',
\eqna\RouseAveragesAppC
$$\eqalignno{
\langle r^a (q^\prime,0) r^b (q,t) \rangle_o
&= {k_B T \over 2 L \epsilon p_q^2} e^{  -{ D \epsilon p_q^2 |t| }  } \,
\delta_{q q^\prime} \delta_{a b}\;,&\RouseAveragesAppC a\cr
\langle \left( r_o^a (t) - r_o^a (0) \right)
           \left( r_o^b (t) - r_o^b (0) \right) \rangle_o
&= {2 D k_B T  \over L} |t| \delta_{a b}\;,&\RouseAveragesAppC b\cr
\langle {\tilde r}^a (q^\prime,0) r^b (q,t) \rangle_o
&=-{i \over 2L} \theta(t) e^{  -{ D \epsilon p_q^2 |t| }  }  \,
\delta_{q q^\prime} \delta_{a b} \;,&\RouseAveragesAppC c\cr
\langle {\tilde r}^a_o (0) r_o^b (t) \rangle_o
&=-{i \over L} \theta(t) \delta_{a b}\;,&\RouseAveragesAppC d\cr
\langle {\tilde r}^a_o (0) {\tilde r}_o^b (t) \rangle_o
&=0\;.&\RouseAveragesAppC e\cr
}$$
Using the above equations we deduce for the segment--correlation function
of the internal modes
\eqnn\RouseAverageSquareAppC
$$\eqalignno{
\langle \left[ r^a (z,t) - r^a (z^\prime,0) \right]&
           \left[ r^b (z,t) - r^b (z^\prime,0) \right] \rangle_o &\cr
= \delta_{a b}
   \Biggl\{ {2 D k_B T \over L} |t| +
            {2 L \over \pi^2} \sum_{q \geq 1} { \T \over \eps q^2 }
            &\left[ {1\over 2} \biggl( \cos(2p_q z) + \cos(2 p_q z^\prime)
\biggr) +
                     \cos(p_q(z+z^\prime)) e^{  -D \epsilon p_q^2 |t| }
           \right] &\cr
+{2 L \over \pi^2} \sum_{q \geq 1} { \T \over \eps q^2 }
            &\left[  1 -   \cos(p_q(z-z^\prime)) e^{  -D \epsilon p_q^2 |t| }
              \right]
   \Biggr\} \;.&\RouseAverageSquareAppC\cr
}$$
This has to be inserted in the corresponding expression for the single line
correlation function
$S^{o} (\k,z,z^\prime,t) = \Gamma^{(2)}_{22} (\k,z,z^\prime,t)$
\eqn\GammatwotwoAppC{
\Gamma^{(2)}_{22} (\k,z,z^\prime,t)  = \langle \rho(-\k,z,t)
\rho(\k,z^\prime,0) \rangle_o
= \exp \left\{
  -{1 \over 4} k^2
  \langle \left[ {\vec r} (z,t) - {\vec r} (z^\prime,0) \right]^2 \rangle_o
         \right\} \;.
}
Since we are dealing with a finite size system there is no translational
invariance with
respect to the $z$-axis, and therefore  correlation functions like
Eq.\GammatwotwoAppC\  depend explicitely on $z$ and $z^\prime$. But, except for
the small region where $z$ and/or $z^\prime$ are close to one of the ends of a
flux line
the second term in Eq.\RouseAverageSquareAppC\ can be neglected. This is due to
the
rapid oscillations of cosine terms like $\cos(2p_qz)$. In the rest of this
appendix
we will neglect those contributions and hence get expressions which depend only
on
the relative coordinate $(z-z^\prime)$.

For $t \geq t_{\rm Rouse} = {L^2 \over D \epsilon}$ the summation
$\sum_{q \geq 1}$ is rapidly converging. If $t \ll t_{\rm Rouse}$ the sums can
be
replaced by integrals
\eqn\SumIntAppC{
{L \over \pi^2} \sum_{q \geq 1} { 1 \over q^2 }
              \left[  1 -   \cos(p_q(z-z^\prime)) e^{  -D \epsilon p_q^2 |t|  }
              \right]
\rightarrow
|z-z^\prime| f \left( {D \epsilon |t| \over (z-z^\prime)^2 } \right) \;,
}
where
\eqn\fFunction{f(y)={1\over\pi}\int_0^\infty{dx\over x^2}\left[1-e^{-y
x^2}\cos x \right]\;.}
The function $f(y)$ can be written in terms of the incomplete
Gamma-function $\Gamma(a,x)$ as
\eqn\faa{f(y) = {1 \over 2} + {1 \over 4 \sqrt{\pi}} \Gamma
\left(-{1 \over 2}, {1 \over 4y} \right) \; , }
with the asymptotic representations
\eqn\fbb{f(y) =
\cases{
\sqrt{{y \over \pi}} \biggl( 1 + { 1 \over 4y } \, ... \, \biggr)
& for $ y \gg 1  \; , $ \cr
{1 \over 2} + {2 y^{3/2} \over \sqrt{\pi}} e^{-1/4y}
\biggl( 1 - 6y \, ... \, \biggr)
& for $ y \ll 1  \; . $}
}
With the latter definitions we find for the two-point function
\eqn\GammatwotwoTransAppC
{\Gamma^{(2)}_{22} (\k,z,t)
= \exp \left[
          -{D \T k^2 \over L} \, |t|
          -  k^2 {\T \over \eps} |z| f \left( { D \epsilon  |t|  \over z^2 }
\right)
          \right]
}
Since $f(0)=1/2$ we recover the static result of Appendix B as we must for
the equal time, $t_i-t_j=0$, correlation function.

Next we consider the single line response function
${\tilde S}^{(0)} (\k,z,z^\prime,t) = \Gamma^{(2)}_{21} (\k,z,z^\prime,t)
= \langle \rho(-\k,z,t) {\tilde \rho} (\k,z^\prime,0) \rangle_o$, again in a
regime where $z$ and $z^\prime$ are not too close to one of the ends
of the flux line. We find
\eqna\GammaResponse$$\eqalignno{
\Gamma^{(2)}_{2 1}(\k,z,t)
\,=\;&iD\T k^2{1\over 2}\langle \rr(0,0)\cdot\r(z,t) \rangle_o &\cr
&\exp \left\{
       -{D \T k^2 \over L} |t|
       - k^2 {\T \over \eps} \,
         |z| f \left( {D \eps |t| \over z^2} \right)
        \right\}&\cr
\,=\;&\theta(t) k^2 {D \T}
    \left[ {1 \over L} + \sqrt{1\over 4 \pi D \epsilon |t|}
e^{-{z^2/(4 D\eps |t|)}} \right]
&\cr
&\times \exp\left\{
-{D \T k^2 \over L} |t|
-k^2{\T\over \eps} |z| f\left({D\eps |t|\over z^2}\right)
\right\},&\GammaResponse a \cr
\Gamma^{(2)}_{1 2}(\k,z,t)\,=\;&\Gamma^{(2)}_{2 1}(\k,z,-t)\;,&\GammaResponse
b\cr
}$$
For the correlation function $\Gamma^{(2)}_{11}(\k,z,t)$ one gets
\eqnn\GammaVanish$$\eqalignno{
\Gamma^{(2)}_{11}(\k,z,t)
=&-(D\T\k)^2 \;
\exp \left\{
 -{D \T k^2 \over L} |t|-
  k^2 {\T \over \eps} \, |z| f \left( {D \eps |t| \over z^2} \right)
     \right\} &\cr
&\times
\Biggl[
 {1\over 2} \langle \rr(z,t)\cdot\rr(0,0) \rangle_o
-{k^2\over 4} \left( \langle \rr(z,t)\cdot\r(z,t) \rangle_o
                    -\langle\rr(z,t)\cdot\r(0,0)\rangle_o
              \right) &\cr
&\phantom{\times {1\over 2} \langle \rr(z,t)\cdot\rr(0,0) \rangle_o -{k^2\over
4} }
 \times \left( \langle\rr(0,0)\cdot\r(z,t)\rangle_o
       -\langle\rr(0,0)\cdot\r(0,0)\rangle_o
 \right)
\Biggr]
\;.\qquad\qquad&\GammaVanish\cr}$$
All the terms inside the square brackets in Eq.\GammaVanish\ \ vanish.
$\langle\rr(z,t)\cdot\rr(0,0)\rangle_o=0$ as can be seen from Eq.\RRG{b} , and
the equal-time correlators $\langle\rr(z,t)\cdot\r(z,t)\rangle_o$ and
$\langle\rr(0,0)\cdot\r(0,0)\rangle_o$ vanish by the causality together with
our choice of causal time discretization, Eq.\causal\ . Finally, the remaining
term
$\langle\rr(z,t)\cdot\r(0,0)\rangle_o\langle\rr(0,0)\cdot\r(z,t)\rangle_o=0$,
because it is proportional to $\theta(t)\theta(-t)$ which vanishes by
the definition of $\theta(t)$ and by the choice of causal time discretization.
Therefore,
\eqn\GammaVanishii{\Gamma^{(2)}_{1 1}(\k,z,t)=0\;.}

We now observe that $\Gamma^{(2)}_{2 2}(\k,z,t)$ and
$\Gamma^{(2)}_{1 2}(\k,z,t)$ satisfy the standard Fluctuation Dissipation
Theorem (FDT),
\eqn\FDT{\pt_t\Gamma^{(2)}_{2 2}(\k,z,t)=
\Gamma^{(2)}_{1 2}(\k,z,t)-
\Gamma^{(2)}_{1 2}(\k,z,-t) \;.}
This is the reason why in the main text we called $\rho_1=\rrho$ the
response field of the physical density field $\rho_2=\rho$ and called
$\Gamma^{(2)}_{2 2}=S^o$ and $\Gamma^{(2)}_{2 1}=\S^o$ the correlation and
response functions, respectively.

Now we discuss the correlation functions in various limiting cases.
For $\T k^2 L/\eps \ll 1$, where the wavelength is much larger than the
static transverse ``radius of gyration'' $\sqrt{2\T L/\eps}$ of the flux line,
the dominant term in
Eq.\GammatwotwoTransAppC\ is the first term. This term describes the center of
mass
motion of a single flux line. We find exponentially decaying correlation
functions
\eqna\LimitcmAppC
$$\eqalignno{
S^o (\k,z,t) \mid_{\T k^2 L/\eps \ll 1}
&\approx  \exp \left[ -{D \T k^2 \over L} |t|  \right] \;,
&\LimitcmAppC a\cr
{\tilde S}^o (\k,z,t) \mid_{\T k^2 L/\eps \ll 1}
&\approx \theta(t){D \T k^2 \over L} \,
\exp \left[-{D \T k^2 \over L} |t|  \right] \;,
&\LimitcmAppC b\cr
}$$
All the other terms are of order $O(k^2 L)$ and can therefore be neglected.

For $\T k^2 L/\eps \gg 1$, where the wavelength is much smaller
than the ``radius of gyration'', the internal modes, i.e. the second term in
Eq.\GammatwotwoTransAppC\ , dominate provided $t$ is not too large.
For times $t \geq t_{\rm Rouse}$ the center of mass motion becomes the
dominant contribution again, since the contribution from the internal modes
scales as $\sqrt{t}$ for large times (note $f(y \gg 1) \sim \sqrt{y}$).  Hence
we
get  for $t\ll t_{\rm Rouse}$
\eqna\LimitintAppC
$$\eqalignno{
S^o (\k,z,t) \mid_{\T k^2 L/\eps \gg 1}
\approx \,
&\exp \left\{
-k^2 \, {\T \over \eps} \, |z| f \left( {D \eps |t| \over z^2} \right)
     \right\}
&\LimitintAppC a\cr
{\tilde S}^o (\k,z,t) \mid_{\T k^2 L/\eps \gg 1}
\approx \, &\theta(t) k^2 \, {\T \over 2} \, \sqrt{D \over \pi \eps t}
\, \exp\left[ {-{z^2 \over 4 D \eps t}}\right] \,&\cr
&\exp \left\{ -k^2 \, {\T \over \eps} \, |z| f \left( {D \eps t \over z^2}
\right)
      \right\}
&\LimitintAppC b\cr
}$$

Finally, we study the Fourier-transforms of these noninteracting structure and
response functions to be used in the main text to calculate the interacting
counterparts. In the center of mass limit one gets a Lorentzian shape
for the correlation function ($f(q=0) = \int_0^L dz f(z)$),
\eqna\LimitFreqcmAppC
$$\eqalignno{
S^o (\k,q=0,\omega) \mid_{\T k^2 L/\eps \ll 1}
&\approx {2 L\Gamma_{\rm cm} (\k) \over \omega^2 + \Gamma_{\rm cm}^2 (\k)} \;,
&\LimitFreqcmAppC a\cr
{\tilde S}^o (\k,q=0,\omega) \mid_{\T k^2 L/\eps \ll 1}
&\approx {L  \over -i \omega +  \Gamma_{\rm cm} (\k)  }
&\LimitFreqcmAppC b\cr
}$$
with the linewidth
\eqn\LinewidthCmAppC
{\Gamma_{\rm cm} (\k)  = {D \T \over L} \, k^2 \;.
}

In the limit where the internal modes dominate we find
\eqn\FT{\S^o(\k,q,\w)=
{\epsilon \over \T} \, {4 \over \sqrt{\pi} k^2}
\int_0^\infty d {\hat t} e^{i {\hat \omega} {\hat t}}
\int_0^\infty dx \cos ( {\hat q} \sqrt{\hat t} x )
e^{-x^2/4 - \sqrt{\hat t}  m(x)} \;,
}
where we have introduced the scaling variables
${\hat \omega} = \omega/\Gamma_\k$, ${\hat q} = 2 q\eps/(\T k^2)$,
and $\Gamma_\k = (\T)^2 D k^4/(4\eps)$,
the noninteracting dynamic relaxation rate.
We have also defined a function $m(x)$ by
\eqn\defm{m(x) = x + x \, {1 \over 2 \sqrt{\pi}} \,
\Gamma \left( -{1 \over 2}, {x^2 \over 4} \right)\;.}
Expanding in powers of ${\hat \omega}$ and we obtain
\eqn\expansionFT{\S^o(\k,q,\w)=
{\epsilon \over \T} \, {4 \over \sqrt{\pi} k^2}
\sum_{n=0}^\infty i^n b_n(\k,q) {\hat \omega}^n}
with the coefficients
\eqn\coeff{b_n(\k,q) = b_n({\hat q}) = 2^{n+1} (2 n + 1) !!
\int_0^\infty dx e^{-x^2/4}
{\cos \left[ (2n+2)
\arctan \left({\hat q} x/m(x)\right) \right] \over
\left[ ({\hat q}x)^2+ m^2(x)  \right]^{n+1}}
}
with the expansions
\eqn\expan{b_n(q,\k) =
\cases{\sim
{\hat q}^{-2n-2}
& for $\hat q \gg 1 \; ,$ \cr
b_n -  {\hat q}^2 b_n^\prime
& for ${\hat q} \ll 1 \; ,$
}
}
where the coefficients are given by
\eqna\constants
$$\eqalignno{
b_n &= 2^{n+1} (2n+1)!! \, \int_0^\infty dx e^{-x^2/4}
m(x)^{-2n-2} \; ,
&\constants a\cr
b_n^\prime &= {(2n+3)!\over n!} \, \int_0^\infty dx e^{-x^2/4} x^2
m(x)^{-2n-4} \; .
&\constants b\cr}
$$

The expression for $S^o(\k,q,\omega)$ can now be easily obtained from the
above expression for $\S^o(\k,q,\omega)$ and the FDT, Eq.\FDT\ . In the limit
of
small frequencies one gets to leading order
\eqna\limits
$$\eqalignno{
\S^o(\k,q,\omega)
&\approx
{\epsilon \over \T} \, {4 \over \sqrt{\pi} k^2}
\left[ b_0({\hat q}) + i b_1({\hat q}) {\hat \omega} -
       b_2({\hat q}) {\hat \omega}^2
\right] \; ,
&\limits a \cr
S^o(\k,q,\omega)
&\approx
{2\over \Gamma_\k}  \, {\epsilon \over \T} \, {4 \over \sqrt{\pi} k^2}
\left[ b_1({\hat q}) -  b_3({\hat q}) {\hat \omega}^2 \right] \cr &
\approx
{2 \over \Gamma_\k}  \, {\epsilon \over \T} \, {4 \over \sqrt{\pi} k^2}
\,
{ {b_1^2 ({\hat q}) / b_3 ({\hat q})}
\over {\hat \omega}^2+ {b_1 ({\hat q})\over b_3 ({\hat q})}}
\; . & \limits b \cr }$$

In the limit ($\w\rightarrow 0$) the Fourier transform $S^o(\k,q,\w)$
can also be written in the Lorentzian form
\eqn\noninterStructii{
S^o(\k,q,\w)
\approx A_s(\k,q)
{2\Gamma_{\rm single} (\k,q) \over \w^2 + \Gamma_{\rm single}(\k,q)^2}
\;,}
where we have introduced the quantities
\eqn\Ak{
A_s(\k,q)  = 2 \, {\epsilon \over \T} \, {4 \over \sqrt{\pi} k^2}
              \, { b_1^{3/2}(\k,q) \over b_3^{1/2}(\k,q) } }
and
\eqn\linewidth{
\Gamma_{\rm single}(\k,q) =
\Gamma_\k \sqrt{b_1(\k,q) \over b_3(\k,q)}\; .}

In the limit, $2 q \epsilon/(\T k^2) \ll 1$, $A_s(\k,q)$ reduces to
\eqnn\Aklimiteins
$$\eqalignno{
A_s(\k,q)
&= 2\, {\epsilon \over \T} \, {4 \over \sqrt{\pi} k^2}
{ b_1 - 6 \left( {q \epsilon \over \T k^2} \right)^2 b_1^\prime
  \over
  b_3 - 2 \left( {q \epsilon \over \T k^2} \right)^2 b_3^\prime
}
&\cr
&\approx 2 {k^2 \T \over \sqrt{\pi} \epsilon} \,
         { b_1 / b_3 \over
           \left( {k^2 \T \over 2 \epsilon} \right)^2 + q^2
           \left[ { 3 b_1^\prime \over 2 b_1 } -
                 {   b_3^\prime \over 2 b_3 } \right]}
&\Aklimiteins\cr }
$$
where $b_n, b_n'$ are constants of order $O(1)$. The formula for
$A_s(\k,q)$ is  quite similar in form to the static structure factor of
non--interacting lines
\eqn\noninterStatics{S^o_s(\k,q)={k^2\T/\eps\over q^2+(k^2\T/2\eps)^2}\;.}
In the same limit the linewidth reads
\eqn\linwlieins{
\Gamma_{\rm single} (\k,q) =
\Gamma_\k \left( {k^2 \T \over 2 \epsilon} \right)^2
{ b_1 / b_3 \over
           \left( {k^2 \T \over 2 \epsilon} \right)^2 + q^2
           \left[ {  b_1^\prime \over 2 b_1 } -
                  {   b_3^\prime \over 2 b_3 } \right]}\; . }

Finally, we discuss the Fourier transform of the single line correlation
function in the limit $q=0$
\eqn\SingleZeroAppC
{S^0(\k,q=0,\omega) =
  \int_0^L dz \int_{-\infty}^\infty dt e^{i \omega t}
  S^0(\k,z,t) \;.
}
Upon introducing the scaling variables
${\tilde \omega} = \omega/(\Gamma_\k (1+{\tilde k}^{-2}))$,
${\tilde k}^2 = k^2 L\T/(4\eps)$, and
$\kappa = \Gamma_\k t_{\rm Rouse}$, where the latter variable gives
the ratio of the Rouse time to the decay time of density fluctuations
we obtain
\eqn\SingleZeroScalingAppC
{S(\k,q=0,\omega) =
{ L  \over \Gamma_\k ( 1 + {\tilde k}^{-2} ) }
  F({\tilde k}, {\tilde \omega}, \kappa)
}
with
\eqnn\Ftilde$$\eqalignno
{&F({\tilde k}, {\tilde \omega}, \kappa) = & \Ftilde\cr
&{2 \over \sqrt{\kappa} }\int_0^\infty d \tau \cos ({\tilde \omega} \tau) \,
 \exp\left[  - {{\tilde k}^{-2} \over 1 +{\tilde k}^{-2} } \tau \right] \,
 \int_0^{\sqrt{\kappa}} dy
 \exp\left[ - \sqrt{\tau \over 1 + {\tilde k}^{-2} }\,
                m \left( y \sqrt{1 + {\tilde k}^{-2} \over \tau}  \right)
\right]&\cr
}$$
The above scaling analysis of the single line correlation function shows that
there are
two time scales. First, there is the time scale for the dynamics of the
internal modes
given by $t_{\rm internal} = 1/\Gamma_{\vec k}$. Then there is a time scale
$t_{\rm Rouse} = L^2/(D \eps)$, which gives the crossover time above which the
dynamics
starts to be dominated by the center of mass mode. Furthermore, we have a
length scale
$R_G^{2D} = \sqrt{2 \T L \over \eps}$ which describes the projected 2D radius
of gyration.

For ${\tilde k} \ll 1$. i.e. for wave length much larger than
the 2D projected Radius of gyration,
the dynamics is determined by the center of mass motion, and we
find for the scaling
function
\eqn\Ftildecm
{F({\tilde k}, {\tilde \omega}, \kappa) =
{ 1 - e^{-\sqrt{\kappa}} \over \sqrt{\kappa}} \,
{2 \over 1 + {\tilde \omega}^2} \; ,
}
which corresponds to Eq.{\LimitFreqcmAppC a}, provided $\kappa$ is small, i.e.
the Rouse time is
small compared to the typical relaxation time for internal modes.

For ${\tilde k} \gg 1$ the internal modes dominate the dynamic structure
factor. In
\fig\FFig{Scaling function $F$ for a fixed value of $\kappa = 1.0$
and a series for ${\tilde k}^{-1}$ ($=0.0001\,{\rm (solid)},\;0.5\,{\rm
(dotted)},\;
1.0\,{\rm (dashed)},\;2.0\,{\rm (dot-dashed)}$).}
the scaling function $F$ is shown for a fixed value of $\kappa = 1.0$
and a series for ${\tilde k}^{-1}$ ($=0.0001,\;0.5,\;1.0,\;2.0$). For any value
of $\kappa$ and ${\tilde k}$ the curves are rather well represented by
\eqn\FtildeAprox
{ F({\tilde k}, {\tilde \omega}, \kappa)
  \approx
  {a_0 \over 1  + a_1 {\tilde \omega}^\gamma}
}
where the coeffiecients $a_0$, $a_1$ and the exponent $\gamma$ depend on
$\tilde k$ and $\kappa$. In the limit of small $\tilde k$ the scaling
function $F$ turns into a Lorentzian with $\gamma = 2$
and $a_0 = 1.264$, $a_1 = 1.0$, which just corresponds
to the Lorentzian shape in the center of
mass limit (compare Eq.{\LimitFreqcmAppC a} and Eq.\Ftildecm).
%
%
%
%
The exponent $\gamma$ decreases with increasing $\tilde k$. The best nonlinear
fit for ${\tilde k}^{-1} = 0.0001$ gives $a_0 \approx 2.93$, $a_1 \approx
5.93$,
and $\gamma \approx 1.344$.
Hence, the lineshape crossover can essentially be characterized in terms of the
effective exponent $\gamma$.
The typical linewidth $\Gamma (k,L)$ of the correlation function is given by
the condition ${\tilde \omega} = 1$, i.e.
\eqn\LinewidthFiniteLAppC
{\Gamma (k,L) = \Gamma_\k \left( 1+ {4 \eps \over \T} \, {1 \over k^2 L}
\right)\;,
}
which describes the crossover from a dynamics determined by the internal modes
to a
dynamics governed by the center of mass motion.

We note that $S^o(\k,z,t)$ and $\S^o(\k,z,t)$ from
Eqs.\GammatwotwoTransAppC\ , \GammaResponse\
satisfy the static limit sum rules, as they must,
\eqna\SumRules$$\eqalignno{S^o_s(\k,q)&=S^o(\k,q,t=0)\;,&\cr
&=\int {d\w\over 2\pi} S^o(\k,q,\w)\;,&\SumRules a\cr
\S^o_s(\k,q)&=\S^o(\k,q,\w=0)\;.&\SumRules b\cr
}$$
\vfill\eject
\vfill\eject
\listrefs
\vfill\eject
\listfigs
\vfill\eject
\bye